\documentclass[10pt, pra, amsmath, onecolumn, showpacs, superscriptaddress]{revtex4-1}
\usepackage{graphicx,color}
 \graphicspath{{./}{./figures/}}
\usepackage{ams math}
\usepackage{enumerate}
\usepackage{mathbbol}
\usepackage{amsfonts}
\usepackage{natbib}

\usepackage{bm}

\usepackage{color}
\usepackage[dvipsnames]{xcolor}

\usepackage[caption=false]{subfig}
\usepackage{bbm}

\newcommand{\nn}{\nonumber}

\newcommand{\sgn}{{\mathrm{sgn}}}

\newcommand{\be}{\begin{equation}}
\newcommand{\ee}{\end{equation}}
\newcommand{\bea}{\begin{eqnarray}}
\newcommand{\eea}{\end{eqnarray}}
\newcommand{\beq}{\begin{eqnarray}}
\newcommand{\eeq}{\end{eqnarray}}

\newcommand{\x}{{\bf x}}
\newcommand{\p}{{\bf p}}
\newcommand{\Ai}{{\rm Ai}}

\newlength{\bilderlength}

\usepackage{subfig}

\begin{document}

\title{Wigner function for noninteracting fermions in hard wall potentials}

\author{Benjamin De Bruyne}
\affiliation{LPTMS, CNRS, Univ. Paris-Sud, Universit\'e Paris-Saclay, 91405 Orsay, France}
\author{David S. Dean}
\affiliation{Univ. Bordeaux and CNRS, Laboratoire Ondes et Mati\`ere  d'Aquitaine
(LOMA), UMR 5798, F-33400 Talence, France}
\author{Pierre Le Doussal}
\affiliation{Laboratoire de Physique de l'Ecole Normale Sup\'erieure, PSL University, CNRS, Sorbonne Universit\'es, 24 rue Lhomond, 75231 Paris, France}
\author{Satya N. \surname{Majumdar}}
\affiliation{LPTMS, CNRS, Univ. Paris-Sud, Universit\'e Paris-Saclay, 91405 Orsay, France}
\author{Gr\'egory \surname{Schehr}}
\affiliation{Sorbonne Universit\'e, Laboratoire de Physique Th\'eorique et Hautes Energies, CNRS UMR 7589, 4 Place Jussieu, 75252 Paris Cedex 05, France}

\date{\today}

\begin{abstract} 
The Wigner function $W_N({\bf x}, {\bf p})$ is a useful quantity to characterize the quantum fluctuations of an $N$-body system in its phase space. 
Here we study $W_N({\bf x}, {\bf p})$ for $N$ noninteracting spinless fermions in a $d$-dimensional spherical hard box of radius $R$ at temperature $T=0$. 
In the large $N$ limit, the local density approximation (LDA) predicts that $W_N({\bf x}, {\bf p}) \approx 1/(2 \pi \hbar)^d$ inside a finite region of the $({\bf x}, {\bf p})$ plane, namely 
for $|{\bf x}| < R$ and $|{\bf p}| < k_F$ where $k_F$ is the Fermi momentum, while $W_N({\bf x}, {\bf p})$ vanishes outside this region, or ``droplet'', on a scale determined by quantum fluctuations. In this paper we investigate systematically, in this quantum region, the structure of the Wigner function along the edge of this droplet, called the Fermi surf. In one dimension, we find that there are three distinct edge regions along the Fermi surf and we compute exactly the associated nontrivial scaling functions in each regime. We also study the momentum distribution $\hat \rho_N(p)$ and find a striking algebraic tail for very large momenta $\hat \rho_N(p) \propto 1/p^4$, well beyond $k_F$, reminiscent of a similar tail found in interacting quantum systems (discussed in the context of Tan's relation). We then generalize these results to higher $d$ and find, remarkably, that the scaling function close to the edge of the box is universal, i.e., independent of the dimension~$d$.   
\end{abstract}

\maketitle

\section{Introduction and main results}

\subsection{Wigner function: overview}

The Wigner function, introduced in quantum mechanics \cite{wigner} and subsequently in the context of signal
processing \cite{ville}, has found a wide variety of applications \cite{case,Bazarov}, ranging from quantum optics \cite{MeasurementWigner,bookQuantumOptics}, trapped atoms and ions \cite{Atom experiment,kur97,Impens,fol14,FermionsWignerMC,Davidson} or electrons in quantum Hall systems \cite{ferraro} all the way to time/frequency analysis \cite{MF1985}. It was initially
introduced to provide a description of quantum mechanics in phase space, i.e. in position and momentum space
$(x,p)$, aiming in particular at a better understanding of the classical limit $\hbar \to 0$ \cite{berry1,Hannay,ADSR1987,CKM1991}. For a single particle in one-dimension, 
described by the wave function $\psi(x)$, the probability density function (PDF) in position space is given by $|\psi(x)|^2$ and 
in momentum space by $|\hat \psi(p)|^2$ where $\hat \psi(p)$ is the Fourier transform of $\psi(x)$. However, because
of the Heisenberg uncertainty principle, it is not possible to simultaneously measure $x$ and $p$. Consequently, one cannot
define, strictly speaking, a {\it joint} PDF of $x$ and $p$ but the closest object to such a joint PDF is the so called Wigner function $W_1(x,p)$,
defined as \cite{wigner}
\bea \label{def_W1}
W_1(x,p) = \frac{1}{2\pi \hbar} \int_{-\infty}^\infty dy \, e^{i p y/\hbar} \psi^*\left( x + \frac{y}{2}\right) \psi \left(x - \frac{y}{2} \right) \;,
\eea
where the subscript '$1$' refers to a single particle. By integrating $W_1(x,p)$ over $p$ (respectively $x$) one can check that one recovers $|\psi(x)|^2$ (respectively $|\hat \psi(p)|^2$). However,
as we will see below, $W_1(x,p)$ is not necessarily positive and for this reason it is sometimes called a ``pseudo'' PDF and, in some cases, the negativity
of the Wigner function has been interpreted as an indicator of non-classicality \cite{KZ2004}.

The Wigner function can also be defined for many-body systems, either bosons or fermions. In particular it has been shown that the 
Wigner function for $N$ fermions trapped in a confining potential, even in the absence of interactions, which will be our main focus here, displays a rich behavior in the limit of 
a large number of fermions $N \gg 1$. This was shown in $d=1$ and at temperature $T=0$ in \cite{balazs,Wiegman} and more recently in any dimension $d \geq 1$ and finite $T$ 
\cite{DDMS2018} for a large class of {\it smooth} confining potentials, such as the harmonic potential. In particular, the behavior of the Wigner function for $N$ particles exhibits, for large $N$, a ``super-universal'' scaling behaviour in the $(x,p)$ plane near the Fermi edge where the Wigner function vanishes. Here the super-universality refers to the fact that the scaling behaviour of the Wigner function is independent of dimension $d$ as well as the shape of the confining potential so long as the potential is smooth~\cite{DDMS2018}. However, much less is known 
about the large $N$ behavior of the Wigner function in the case of {\it non-smooth} or {\it singular} potentials, such as the hard box potential (see however \cite{ADSR1987,CKM1991,AD1992,BDR2004,LS2013} mainly in the nuclear physics literature). In this paper, we show that this case also displays very rich behaviors, which are however markedly different from the one found for smooth potentials.

Let us consider $N$ noninteracting fermions in $d$ dimensions and in the presence of a trapping potential $V(\hat {\bf x})$. 
The many-body Hamiltonian is
${\cal H}_N = \sum_{i=1}^N \hat H(\hat {\bf x}_i, \hat {\bf p}_i)$ expressed in terms of the single particle Hamiltonian
\bea \label{def_H}
\hat H(\hat {\bf x}, \hat {\bf p}) = \frac{\hat {\bf p}^2}{2m} + V(\hat {\bf x}) \;,
\eea    
where $m$ is the mass of the fermions. During the last few years, trapped Fermi gases have generated tremendous interest, both theoretically \cite{GPS08,Kohn,Eis2013,us_finiteT,DPMS:2015,fermions_review,marino_prl,lacroix_EPL,lacroix2018non,CLM2015,Dub2017, manas2018, jm2019} and 
experimentally in cold atom systems \cite{BDZ08,Cheuk2015,Parsons2015,Muk2017,Hol2021}. From the theoretical point of view, the case of $d=1$ is particularly interesting
since, for some specific potentials $V(x)$, the positions of the fermions in the ground state of ${\cal H}_N$ can be mapped to the eigenvalues of
certain ensembles of random matrices (for a recent review see \cite{DDMS2019}). For instance, the case of the harmonic potential $V(x) = m \omega^2 x^2/2$ corresponds to the Gaussian Unitary Ensemble (GUE) \cite{Eis2013,marino_prl}, while the
hard box potential, i.e., $V(x) = 0$ if $|x| \leq R$ and $V(x) = + \infty$ elsewhere, corresponds to the Jacobi Unitary Ensemble (JUE) of random matrices~\cite{lacroix_EPL,lacroix2018non}. These 
ensembles are well known
in random matrix theory to display rather different behaviours \cite{mehta,forrester}. In both cases, the spatial density of fermions, for large $N$, has a finite support, i.e., the density vanishes beyond a certain 
value, for $|x| > x_{\rm edge}$, which defines an edge in the $x$-space. Of course for the hard box, $x_{\rm edge} = R$ -- this is called a {\it hard edge} in RMT -- while for the harmonic oscillator $x_{\rm edge} = \sqrt{2N}/\alpha$ where $\alpha = \sqrt{m \omega/\hbar}$ is the inverse oscillator length -- this is referred to as a {\it soft edge} in RMT. Similarly, one expects that the density in momentum space also exhibits an edge at some value $p_{\rm edge}$. In the case of the harmonic potential the positions and momenta have the same statistics, in particular at the edge. The more general case of a smooth potential $V(x) \sim x^{2n}$ with $n\geq 1$ and integer was studied in \cite{DMS2018} and it was shown that there exist several different universality classes indexed by $n$, which have also generated some interest in the mathematical physics literature \cite{cafasso,BBW2020,KZ2020}. However, the case of hard box potentials (which would formally correspond to the limit $n \to \infty$) has not been studied so far. Nevertheless it is natural to expect that they behave differently from the case of a smooth potential. This strongly suggests that the Wigner function of a hard box potential, not only in $d=1$ but also in higher dimensions $d>1$,  
will display a large $N$ behavior in phase space $(\x,\p)$, where ${\bf x} = (x_1, x_2, \cdots, x_d)$ and similarly ${\bf p} = (p_1, p_2, \cdots, p_d)$, that will be different from the behavior found for smooth potentials \cite{DDMS2018}.

We focus on the zero temperature limit where the system is described by the many-body ground state wave-function $\Psi_0({\bf x}_1, {\bf x}_2, \ldots, {\bf x}_N)$. The many-body Wigner function, i.e., the  
generalization of the formula (\ref{def_W1}) to any $N$ and $d$ is given by \cite{wigner}
\bea \label{def_WN}
W_N({\bf x}, {\bf p}) = \frac{N}{(2 \pi \hbar)^d} \int_{-\infty}^\infty d{\bf y} d{\bf x}_2 \ldots d{\bf x_N} \, e^{i {\bf p}\cdot {\bf y}} \Psi_0^*\left({\bf x}+ \frac{{\bf y}}{2}, {\bf x}_2, \ldots, {\bf x}_N\right)\Psi_0\left({\bf x}- \frac{{\bf y}}{2}, {\bf x}_2, \ldots, {\bf x}_N \right) \;.
\eea
One can easily check that $W_N({\bf x}, {\bf p})$ in (\ref{def_WN}) satisfies the relations
\bea \label{marginals}
\int_{-\infty}^\infty d\p \, W_N({\bf x},{\bf p}) = \rho_N({\x}) \quad, \quad \int_{-\infty}^\infty d\x \, W_N({\bf x},{\bf p}) = \hat \rho_N({\p}) \quad {\rm and} \quad \int_{-\infty}^\infty \,d\x\,d\p \, W_N({\bf x},{\bf p}) = N \;,
\eea 
where $\rho_N(\x)$ and $\hat \rho_N({\bf p})$ denote respectively the density in real and momentum space in the ground state, i.e.,
\bea\label{def_density}
\rho_N(\x) = \sum_{i=1}^N \left< \delta(\x - \x_i) \right >_0 \quad, \quad \rho_N(\p) = \sum_{i=1}^N \left< \delta(\p - \p_i) \right >_0  \;,
\eea
which are both normalized to $N$, i.e. $\int_{-\infty}^\infty d{\bf x} \, \rho_N(\x) = \int_{-\infty}^\infty d{\bf p}\, \hat \rho_N(\p) = N$. In (\ref{def_density}) the notation $\langle \ldots \rangle_0$ denotes an average computed in the many-body ground-state $\Psi_0$.  

The many-body Wigner function in Eq. (\ref{def_WN}) is seemingly a complicated object for finite $N$ as it depends on the details of the trapping potential $V(\bf x)$. Remarkably however it turns out
that in the limit of large $N$ the Wigner function $W_N({\bf x}, {\bf p})$ reaches a rather simple limiting form which is universal. Indeed, in the limit $N \to \infty$, $W_N({\x},\p) \simeq 0$ outside a domain $\Gamma$, which is just the region of the space phase $(\x, \p)$ that is allowed classically. Inside this region $\Gamma$ the Wigner function is a constant, i.e., $W_N(\x,\p) \simeq 1/(2 \pi \hbar)^{d}$. One can indeed show that, as $N \to \infty$, the expression in (\ref{def_WN}) takes the very simple form 
\bea \label{LDA}
W_N(\x, \p) \simeq \frac{1}{(2\pi \hbar)^d} \Theta(\mu - E(\x,\p))
\eea  
where $\mu$ is the Fermi energy and 
\bea\label{def_E}
E(\x,\p) = \frac{\p^2}{2m} + V(\x) \;
\eea
is the classical energy of a single particle. In Eq. (\ref{LDA}), $\Theta(z)$ is the Heaviside step function such that $\Theta(z)=1$ if $z>0$ and $0$ otherwise. While the result in (\ref{LDA}) can be obtained via semi-classical methods, such as the local density approximation \cite{castin}, it was also derived in  \cite{DDMS2018} via a controlled asymptotic analysis of the exact formula in (\ref{def_WN}). It is clear from (\ref{LDA}) that $W_N(\x,\p)$ vanishes outside the domain $\Gamma$ -- sometimes called a ``droplet'' -- delimited by the surface $(\x_e, \p_e)$ described by
\bea \label{Fermi_surf_0}
\frac{\p_e^2}{2m} + V(\x_e) = \mu \;,
\eea 
which, following Ref. \cite{Wiegman}, we will call the ``Fermi surf''. In Eq. (\ref{Fermi_surf_0}), the subscript '$e$' refers to the {\it edge} region of the droplet $\Gamma$, i.e. the vicinity of the Fermi surf, as opposed to the {\it bulk} region, far from the Fermi surf.

In fact, the simple form in (\ref{LDA}) holds only for $\x$ and $\p$ in the bulk, i.e. for $\x$ and $\p$ far enough from the Fermi surf (\ref{Fermi_surf_0}). At the edge, i.e. close to $(\x_e, \p_e)$, one expects that the sharp step function will be smoothened over a certain energy scale $e_N$. It is then natural to ask how this scale $e_N$ together with the precise form of the Wigner function close to the Fermi surf (\ref{Fermi_surf_0}) depend on the trapping potential $V(\x)$ and on the space dimension $d$. In Ref. \cite{DDMS2018} this question was addressed for the wide class of {\it smooth} confining potentials, i.e., for potentials that behave for large $|\x|$ as $V(\x) \propto |\x|^m$ (for some real number $m>0$). In this case, it was demonstrated that, in terms of the dimensionless variable $a$ defined as
\bea \label{def_a}
a = \frac{1}{e_N} (E({\bf x},{\bf p}) -  \mu)\;, 
\eea
where $(\x, \p)$ is a point in the phase space close to the Fermi surf and $e_N$ is an energy scale given by
\beq
e_N = \frac{(\hbar)^{2/3}}{(2 m)^{1/3}} \left( \frac{1}{m} (\p_e \cdot  \nabla)^2 V({\bf x}_e)  + |\nabla V({\bf x}_e)|^2 \right)^{1/3} \;, \label{eN_intro} 
\eeq 
the Wigner function $W_N(\x,\p)$ takes the scaling form
\be
W_N({\bf x},\p)  \simeq \frac{{\cal W}(a)}{(2 \pi \hbar)^d}  \label{scal1} \;.
\ee
Quite remarkably, it was shown in \cite{DDMS2018} that the scaling function ${\cal W}(a)$ is ``super-universal'', i.e., independent of both the potential and the space dimension $d$ and is given by \cite{DDMS2018}
\be
{\cal W}(a) = \int_{2^{2/3} a}^{+\infty} \Ai(u) du \label{scal2}  \;,
\ee
where ${\Ai}(u)$ is the Airy function. The function ${\cal W}(a)$ has the asymptotic behaviors
\begin{eqnarray}
{\cal W}(a) \sim
\begin{cases}\label{asymp_plus}
& 1\;, \;  \hspace*{4.3cm} a\to -\infty \\
& \\
&(8\pi)^{-1/2}\, a^{-3/4}\,
\exp\left[-\frac{4}{3}\,
a^{3/2}\right]\;, \; a\to + \infty  \;.
\end{cases}
\end{eqnarray}
In particular, the limit $\lim_{a \to -\infty} {\cal W}(a)=1$ ensures
a smooth matching with the bulk result \eqref{LDA}. 

\subsection{Model and main results} 

The goal of this paper is to investigate the case when the potential $V(\x)$ is a non-smooth function. In particular, we consider the case where $V(\x)$ is a hard box potential, i.e., 
     \begin{align}\label{def_V_Box}
     V(\mathbf{x})=
     \begin{cases}
       0, \quad &|\mathbf{x}|\leq R \\
       \infty, \quad& |\mathbf{x}|> R \;.
     \end{cases}\, 
   \end{align}
It is useful to summarise our main results. For simplicity, in the remaining of this section we set $m = \hbar= R =1$.    

\begin{figure}[t]
    \centering
    \includegraphics[width=0.4\textwidth]{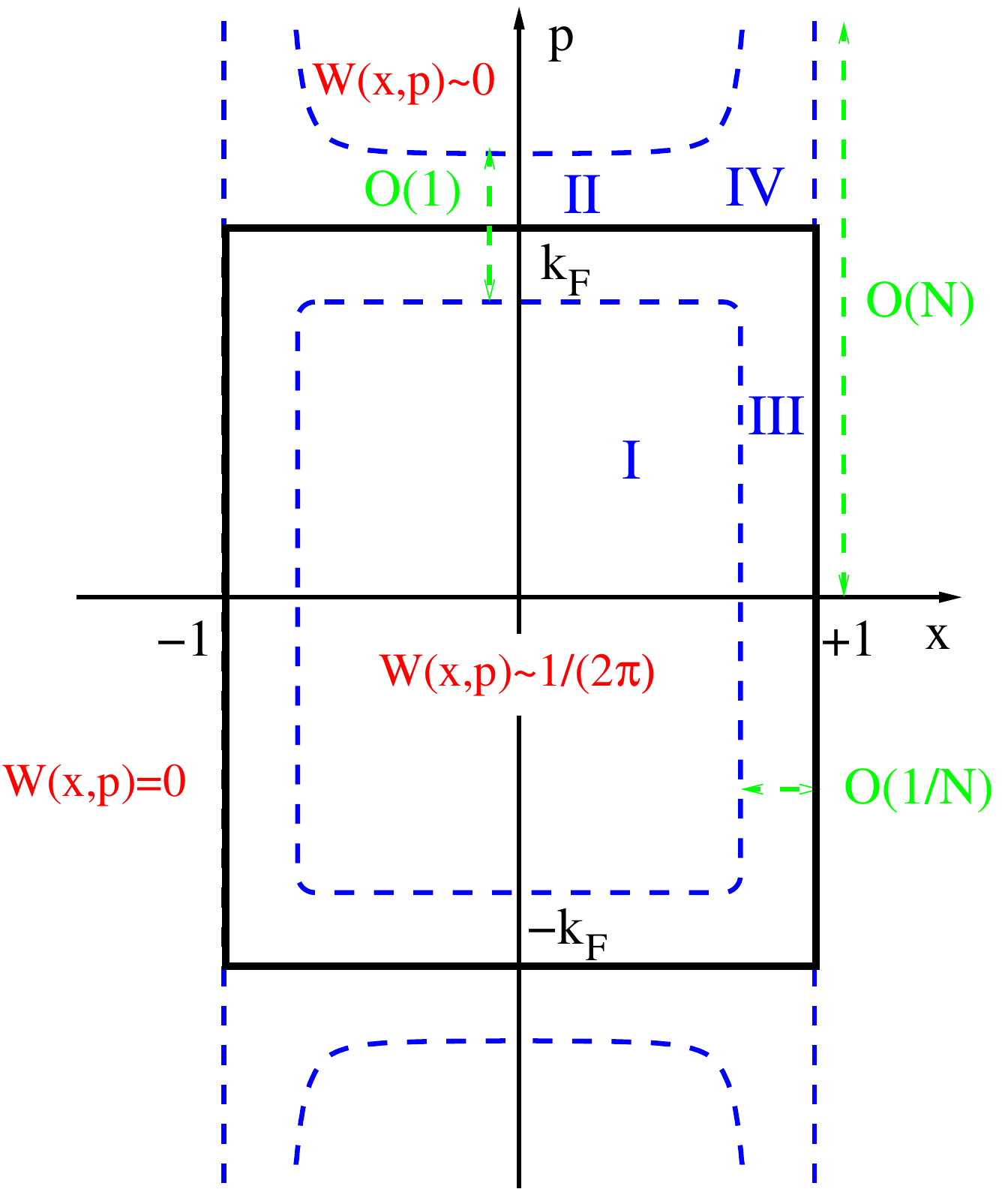}
    \caption{Representation in phase space $(x,p)$ of the various regimes for the Wigner function for the $1d$ hard box in the limit of a large number of fermions. The thick black rectangle is the Fermi surf (with $k_F = N \pi/2$). The inside of this
    rectangle is the bulk region (I) where the Wigner function is approximately constant and non zero. Outside of this region it is approximately zero (and strictly zero for $|x|>1$). 
    The regions where the various crossovers studied in the text take place, i.e., near the Fermi momentum (II), near the hard wall (III), and near the corner (IV) are indicated. The region III extends over a momentum scale of order $O(k_F) = O(N)$, while the region II extends only to order $O(1)$ in momentum space. The blue dashed lines represent schematically the lines of constant value of the Wigner function. }
    \label{fig:fermisurf}
\end{figure}
\vspace*{0.5cm}

\subsubsection{One dimension $d=1$}

\begin{figure}[htb]
\subfloat[Wigner function $W_N(x,p)$ for $N=20$ fermions (\ref{Wigner_alter}).]{%
\includegraphics[width=0.6\textwidth]{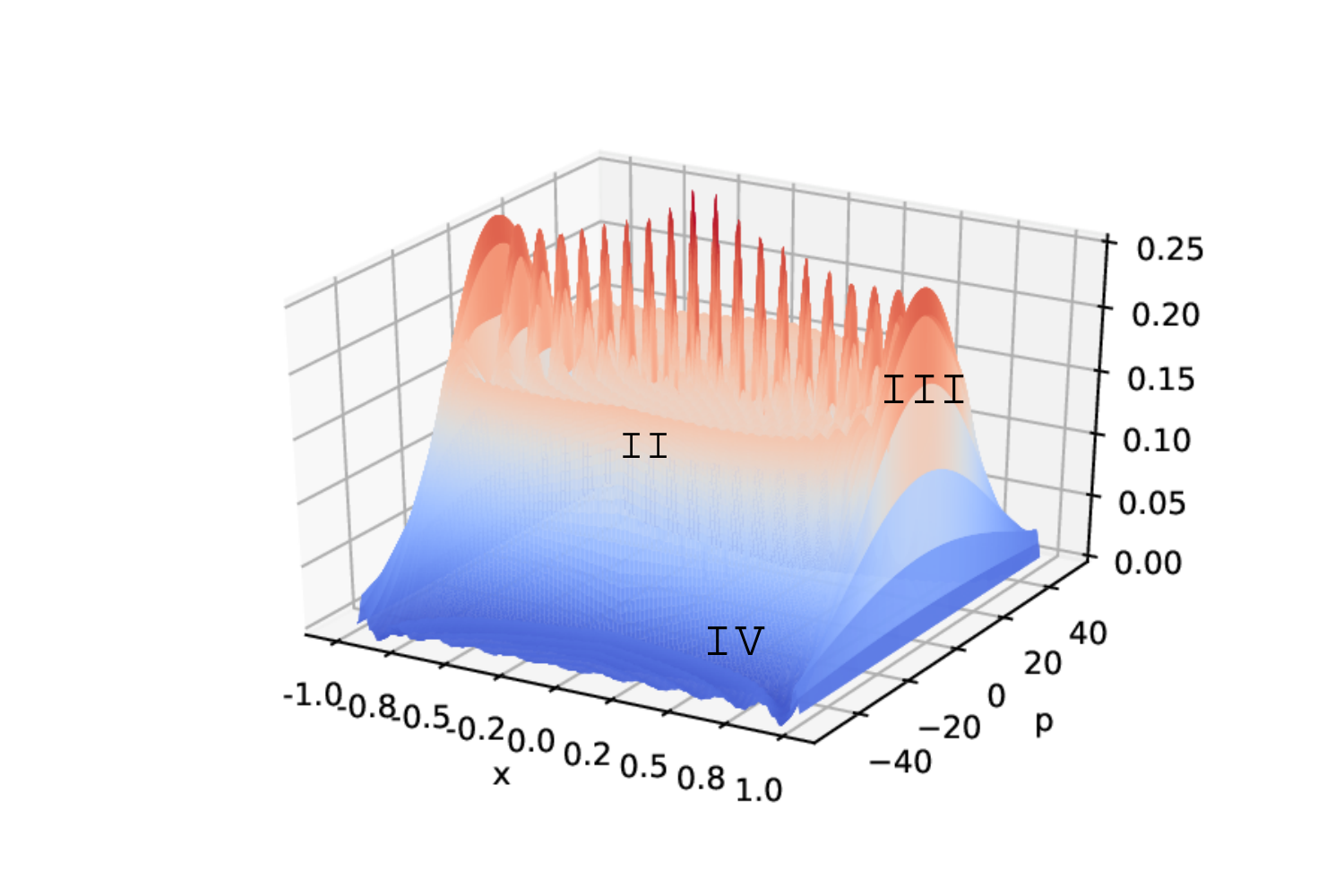}%
}\\
\subfloat[ Scaling function in region II (\ref{region_II}):
          ${\cal W}_{\rm II}\left(x,q \right)\simeq W_N\left(x,p= \frac{\pi}{2}(N+q)\right)$.]{%
\includegraphics[width=0.32\textwidth]{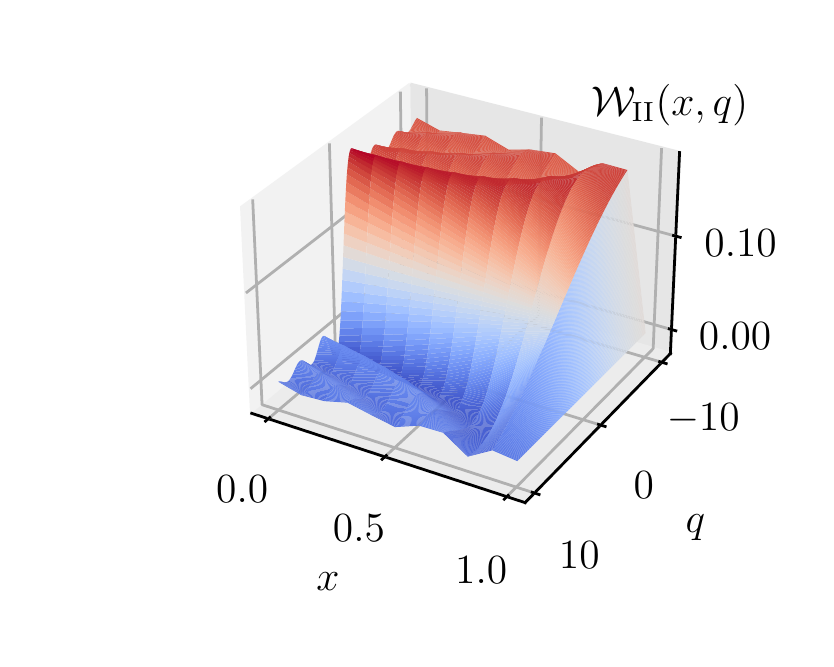}%
}\hfill
\subfloat[Scaling function in region III (\ref{region_III}):
          ${\cal W}_{\rm III}\left(\tilde s,\tilde p \right)\simeq  W_N\left(x = 1 -  \tilde s/k_F, p = \tilde p \, k_F\right)$.]{%
\includegraphics[width=0.32\textwidth]{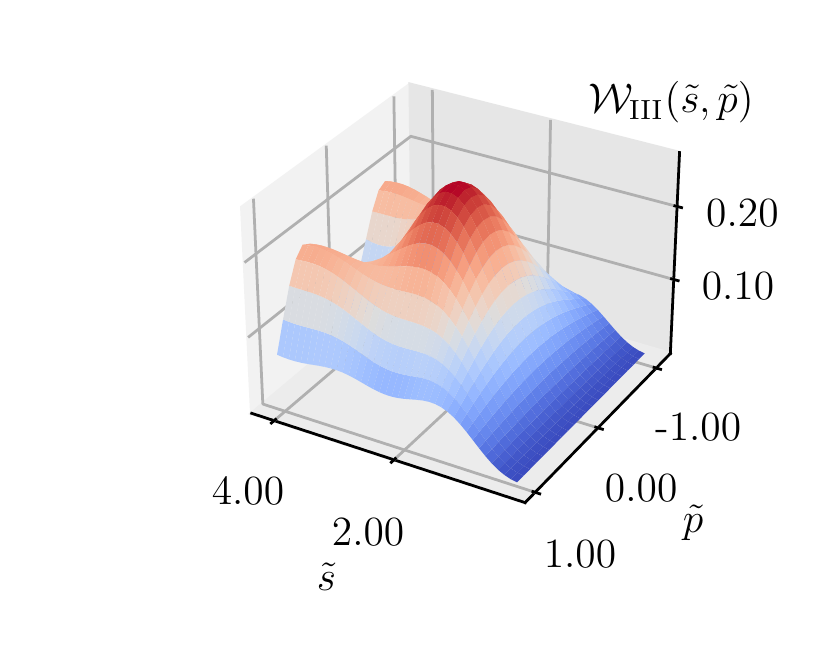}%
}\hfill
\subfloat[Scaling function in region IV (\ref{region_IV}):
          ${\cal W}_{\rm IV}\left((1-x)(p-k_F)\right)\simeq  W_N\left(x,p\right) $.]{%
  \includegraphics[width=0.32\textwidth]{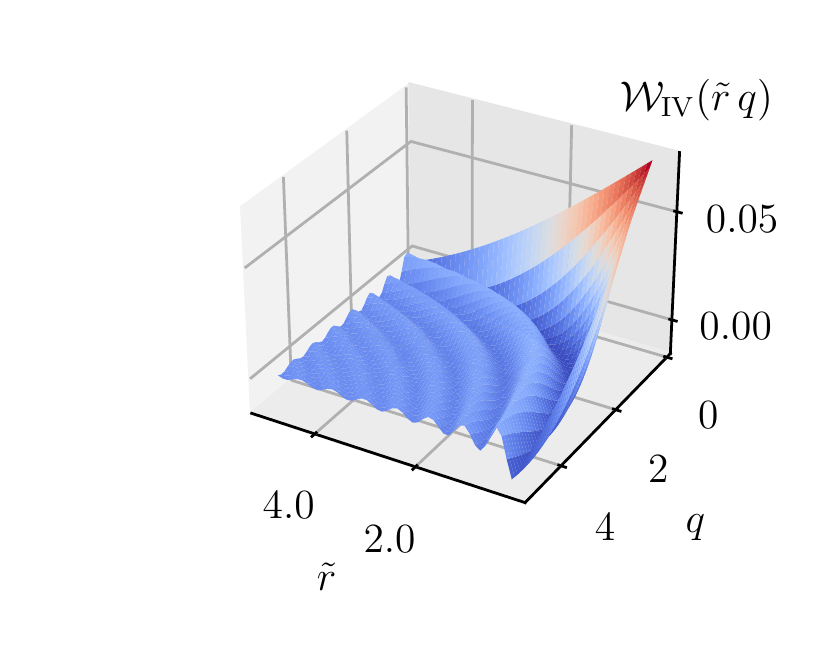}%
}\hfill
\caption{\textbf{a):} Exact Wigner function $W_N(x,p)$ for $N=20$ fermions. In the large $N$ limit, the Wigner function exhibits scaling regimes in different regions of the border of the Fermi surf (see figure \ref{fig:fermisurf}). The location of the scaling regimes are labelled by the roman numerals ${\rm II}$, ${\rm III}$ and ${\rm IV}$ and closer views of the scaling regimes are displayed in the lower panel using rescaled coordinates. \textbf{b):} Scaling function ${\cal W}_{\text{II}}(x,q)$  at the momentum edge of the Fermi surf (region II in figure~\ref{fig:fermisurf}).  \textbf{c):} Scaling function ${\cal W}_{\text{III}}(\tilde s,\tilde p)$  near the wall of the hard box (region III in figure \ref{fig:fermisurf}). \textbf{d):} Scaling function ${\cal W}_{\text{IV}}(\tilde r\, q)$  in the corner of the Fermi surf (region IV in figure \ref{fig:fermisurf}). The scaling function ${\cal W}_{\rm IV}(z)$ only depends on one variable but for illustrative purposes, we used the rescaled coordinates $x=1-\frac{\tilde r}{k_F^\alpha}$ and $p=k_F +q k_F^\alpha $ with $\alpha<1$.}
  \label{fig:3dplots}
\end{figure}

{\it Wigner function.} In this case, the Wigner function $W_N(x,p)$ for the hard box (\ref{def_V_Box}) can  be computed exactly for any value of $N$ [see Eq. (\ref{eq:Wua}) and Fig. \ref{fig:3dplots}]. 
From the Wigner function, we also obtain the exact expression for the density in position (\ref{density_x}) and momentum (\ref{density_p}) space (see also Fig. \ref{fig:density}). 
From Eq. (\ref{LDA}), setting $V(x)= 0$, we immediately see that the Fermi surf is very simple in this case, and given by the rectangle passing through the four corners (see Fig. \ref{fig:fermisurf})
\bea \label{Fermi_surf}
(-1,-k_F), (1,-k_F), (1,k_F), (-1,k_F) \quad {\rm with} \quad k_F = \sqrt{2 \mu}\quad, \quad {\rm in \; the} \; (x,p) \; {\rm plane} \;,
\eea
where $k_F$ is the Fermi wave vector. Outside this rectangle, and far enough from the Fermi surf, $W_N(x,p) \simeq 0$ in the limit $N \to \infty$. We find that the rest of the $(x,p)$ plane is divided in four regions (see Fig. \ref{fig:fermisurf}), one bulk region (I) well inside the Fermi surf and three edge regions (II, III, IV and their symmetric counterparts) close to the Fermi surf where the Wigner function exhibits different scaling regimes in the limit $N \to \infty$: 
\begin{itemize}
\item[$\bullet$]{(I) For $-1<x<1$ and $-k_F < p < k_F$ the Wigner function is constant, i.e., 
\bea \label{region_I}
W_N(x,p) \simeq \frac{1}{2 \pi} {\cal W}_{\rm I}(x,p) \quad {\rm with} \quad {{\cal W}_{\rm I}(x,p) = 1} \;, \; N \to \infty \;,
\eea}
in agreement with the LDA prediction (\ref{LDA}).
\item[$\bullet$]{(II) For $-1<x<1$ and $p$ close to the momentum edge $(|p| - p_e) = O(1)$, the Wigner function takes the scaling form for large $N$, say for $p$ close to $p_e=+k_F$,  
\bea \label{region_II}
W_N(x,p) \simeq \frac{1}{2 \pi} {\cal W}_{\rm II}\left(x,\frac{2}{\pi}(p-k_F)\right) \quad, \quad  \mathcal{W}_{\text{II}}(x,q) = \frac{1}{\pi} \sum_{m=0}^\infty \frac{\sin((m+q) \pi (1-x))}{m+q} \;.
\eea
A plot of the scaling function ${\cal W}_{\rm II}(x,q)$ is shown in Fig. \ref{fig:3dplots} b) and Fig. \ref{fig:edgemom}, while its asymptotic behaviors are given in Eq. (\ref{largek_II}) and (\ref{top_right}). 
} 
\item[$\bullet$]{(III) For $x$ close to the (right) wall $(1-x) = O(1/k_F) = O(1/N)$ and $p = O(k_F) = O(N)$, we find that $W_N(x,p)$ takes the scaling form for large $N$
\bea \label{region_III}
W_N(x,p) \simeq \frac{1}{2 \pi} {\cal W}_{\rm III}(k_F(1-x), p/k_F) \;, \;  {\cal W}_{\rm III}(\tilde s, \tilde p) = \frac{\text{Si}(2(1+\tilde p)\tilde s)+\text{Si}(2(1-\tilde p)\tilde s)}{\pi} - \frac{\sin(2\tilde s)\sin(2\tilde p \tilde s)}{\pi \tilde p \, \tilde s}\,,
\eea
where Si is the sine integral function $\text{Si}(x)=\int_0^x \sin(t)/t\, dt$. A plot of the function ${\cal W}_{\rm III}(\tilde s, \tilde p)$ is shown in Fig. \ref{fig:3dplots} c) and Fig. \ref{fig:wall}, while its asymptotic behaviours are given in Eqs. (\ref{W3_smalls}), (\ref{W3_larges}) and (\ref{W3_largep}).

}
\item[$\bullet$]{(IV) Finally, for $x$ and $p$ near the top right corner region, we identify a scaling region of ``mesoscopic'' size with $k_F^{-1} \ll 1-x \ll 1$ and $1\ll |p - k_F| \ll k_F$ but keeping the product $(1-x)(p-k_F) = z$ fixed (of course, a similar scaling holds near the other three corners of the Fermi surf). 
In this regime one finds that $W_N(x,p)$ takes the scaling form
\bea \label{region_IV}
W_N(x,p) \simeq \frac{1}{2\pi} {\cal W}_{\rm IV}\left((1-x)(p-k_F)\right) \;, \;  {\cal W}_{\rm IV} (z)=\frac{1}{2} - \frac{\text{Si}(2\,z )}{\pi} \;.
\eea
As discussed below, this regime smoothly interpolates between the regime II (where $(p-k_F) = O(1)$) and the regime III (where $1-x = O(k_F^{-1})$). A plot of ${\cal W}_{\rm IV}(z)$ is shown in Fig. \ref{fig:3dplots} d), while its asymptotic behaviors are given in Eq. (\ref{W4asympt}). 
}
\end{itemize}

\begin{figure}[t]
    \centering
    \includegraphics[width=0.4\textwidth]{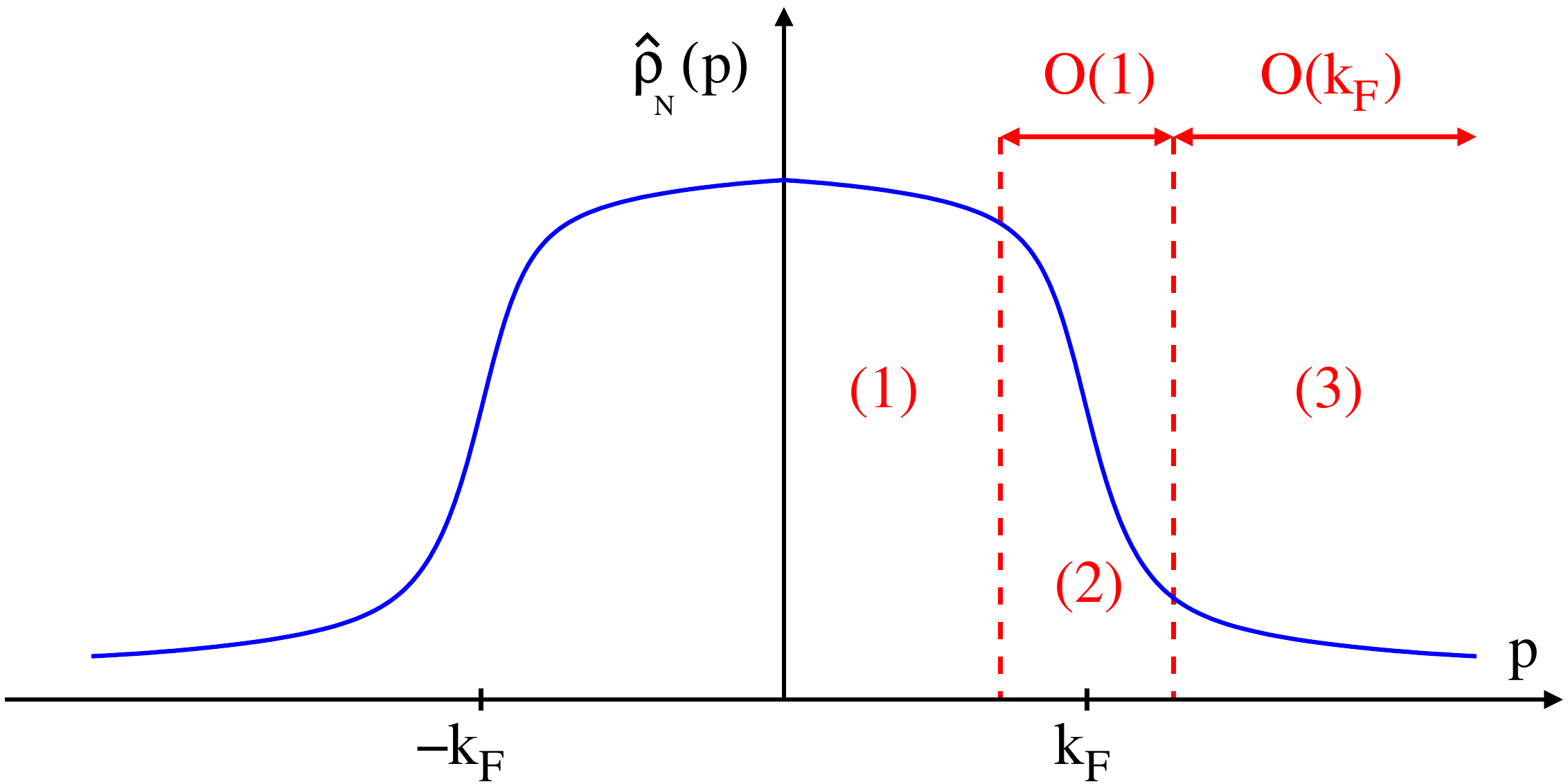}
    \caption{Sketch of the momentum density $\hat \rho_N(p)$ vs $p$ for the $1d$ hard box in the limit of a large number of fermions. In the region (1), for $|p| < k_F$ the density is approximately constant $\hat \rho_N(p) \simeq 1/\pi$. Outside of this bulk region, we find that there are two distinct edge regimes: the regime (2), i.e., near the Fermi surf with $(p-k_F) = O(1)$, corresponds to fermions which are inside the box, i.e. far from the wall, and a far tail regime (3) where $p = O(k_F)$, which corresponds to fermions which are close to the wall. In the latter regime (3) the density has an algebraic tail $\hat \rho_N(p) \propto p^{-4}$ for $|p| \gg k_F$}
    \label{fig:rhos}
\end{figure}

{\it Mean density in real space and momentum space}. It is also interesting to analyse separately the large $N$ behavior of the densities, both $\rho_N(x)$ in $x$-space and $\hat \rho_N(p)$ in $p$-space. The analysis of $\rho_N(x)$
was recently carried out in Refs. \cite{lacroix_EPL,lacroix2018non}. In the bulk, for $-1<x<1$, the density $\rho_N(x)$ can be easily obtained by integrating the Wigner function $W_N(x,p)$ in region I in Eq. \eqref{region_I} [see Eq. \eqref{marginals}]
\bea \label{rhox_1}
\rho_N(x) = \int W_N(x,p) \, dp \simeq \frac{1}{2\pi} \int_{-k_F}^{k_F} \, dp = \frac{k_F}{\pi} \;,
\eea
i.e., the density is, as expected, uniform in the bulk, i.e. far from the wall [see Fig. \ref{fig:density} a)]. On the other hand, close to wall, the density vanishes over a scale $1/k_F$ and is described by the scaling form \cite{lacroix2018non}
\bea \label{rhox_2}
\rho_N(x) \simeq \frac{k_F}{\pi} F_1(k_F(1-x)) \quad, \quad F_1(\tilde s) = 1 - \frac{\sin{2 \tilde s}}{2\tilde s} \;.
\eea
There are thus two regimes for the density: (i) the bulk for $-1<x<1$ [see Eq. \eqref{rhox_1}] and (ii) the edge of the box, near the wall for $1-x = O(1/k_F)$ [see Eq. \eqref{rhox_2}] . We show here that the density in $p$-space exhibits three different regimes (see Fig. \ref{fig:rhos}): 
\begin{itemize}
\item[$\bullet$] (1) For $-k_F<p<k_F$, the density $\hat \rho_N(p)$ can be obtained by integrating the Wigner function $W_N(x,p)$ given, in region I, in Eq. \eqref{region_I} [see Eq. \eqref{marginals}]
\bea \label{rhop_1}
\hat \rho_N(p) = \int W_N(x,p) \, dx \simeq \frac{1}{2\pi} \int_{-1}^{1} \, dx = \frac{1}{\pi} \;,
\eea
i.e., the momentum density is also uniform, as in position space in the bulk (\ref{rhox_1}), [see also Fig. \ref{fig:density} b)].
\item[$\bullet$] (2) For $p$ close to $k_F$, with $p-k_F = O(1)$, the density takes a nontrivial limiting form
\bea \label{rhop_2}
\hat \rho_N(p) \simeq \frac{1}{\pi} \hat F_1\left(\frac{2}{\pi}(p-k_F)\right) \quad, \quad \hat F_1(q) =  \frac{1}{4\pi ^2}\left(4 \psi ^{(1)}( q)+\cos (\pi  q) \left[\psi ^{(1)}\left(\frac{ q+1}{2}\right)-\psi
   ^{(1)}\left(\frac{ q}{2}\right)\right]\right)\,,
\eea
where $\psi ^{(1)}(z) = \sum_{k=0}^\infty 1/(k+z)^2$ is the tri-gamma function. The asymptotic behaviors of this function are given in Eq. (\ref{eq:tailrhop}) while a plot of $\hat F_1(q)$ is shown in Fig.~\ref{fig:edgena}.

\item[$\bullet$] (3) For $p = \tilde p \, k_F$ with $\tilde p>1$, we find yet another nontrivial regime where the density $\hat \rho_N(p)$ takes the scaling form
\bea \label{rhop_3}
\hat \rho_N(p) \simeq \frac{1}{k_F} \hat {\sf F}_1\left(\tilde p=\frac{p}{k_F}\right) \quad, \quad  \hat {\sf F}_1(\tilde p)=\frac{\tilde p+(1-\tilde p^2)\coth ^{-1}\tilde p}{\pi ^2  \tilde p \left(\tilde p^2-1\right)}\;, \; \tilde p > 1 \;.
\eea
The asymptotic behaviors of this function are given in Eq. (\ref{asympt_rhop_3}) and a schematic plot of the various regimes for $\hat {\sf F}_1(\tilde p)$ is presented in~Fig.~\ref{fig:edgena2}. Note in particular that, for large $\tilde p$, the momentum density has an algebraic tail $\hat {\sf F}_1(\tilde p) \sim 1/\tilde p^4$. In fact, we obtain a more precise formula, valid for any finite $N$ for this momentum tail distribution  
\bea \label{tail_rhoN_intro}
\hat \rho_N(p) \simeq \frac{\pi}{96 p^4} N (N+1)(2N+1) \simeq \frac{E_N}{2\pi} \frac{1}{p^4} \underset{N \to \infty}{\simeq} \frac{2}{3 \pi^2} \frac{k_F^3}{p^4}  \;,
\eea
where $k_F = N\pi/2$ and $E_N$ is the ground state energy. Interestingly, the same algebraic tail $\sim 1/\tilde p^4$ also appears for particles (bosons or fermions) interacting via a contact repulsion, where it is known under the name of Tan's relations \cite{MVT2002,VGPS2004,Tan2008,BZ2011, BD2021}. Remarkably, we also find a $1/\tilde p^4$ decay even for noninteracting fermions but in the presence of a hard box potential. This can be interpreted as a consequence of the effective repulsion between a fermion and its image across the hard wall.

\end{itemize}

In summary we find that the structure of the momentum distribution is quite rich.
Indeed we note that the tail behavior of the momentum distribution in the case of a hard box potential has a markedly different behavior from that of a smooth confining potential. In the hard box case, we have two distinct edge regimes in the momentum space (see Fig. \ref{fig:rhos}): the regime (2), i.e., near the Fermi surf with $(p-k_F) = O(1)$, corresponds to fermions which are inside the box, i.e. far from the wall,
and a far tail regime (3) where $p = O(k_F)$, which corresponds to fermions which are close to the wall. We also note that, when $\hat \rho_N(p)$ is integrated over these two tail regions (2) and (3), it contributes to order $O(1)$, indicating that this corresponds to a single outlier with an extremely high momentum. Finally, we note that the $1/p^4$ tail for the momentum distribution, found here in the presence of a hard wall, is very different from the far tail behavior of the momentum distribution in a smooth confining potential, such as the harmonic well where $\hat \rho_N(p)$ decays faster than an exponential \cite{DMS2018}.   
\\

{\it Kernel.} To quantify the quantum correlations beyond the density and the Wigner function, it is useful to calculate higher order correlation functions of the fermion positions $x_i$
and momenta $p_i$. In the ground state the positions $x_i$'s form a determinantal point process (DPP), and similarly for the momenta $p_i$'s. A central building block for DPPs is the so called kernel 
$K_N(x,x')$ (in position space) or  $\hat K_N(p,p')$ (in momentum space). Any $n$-point correlation function, either in position or in momentum space, can be expressed as an $n \times n$ determinant
whose entries are given by the kernel. Indeed, the Wigner function $W_N(x,p)$ discussed so far can also be expressed in terms of the kernel by the relation \cite{DDMS2018}
\bea \label{W_vs_Kernel}
W_N(x,p) = \frac{1}{2 \pi} \int dy \, e^{i p y} \, K_N\left(x-\frac{y}{2}, x + \frac{y}{2} \right) \;.
\eea
For the hard wall potential, the kernel in real space $K_N(x,x')$ 
was studied in detail in Ref. \cite{lacroix2018non}. 
Here we compute the kernel in momentum space $\hat K_N(p,p')$ for the hard wall case. In the region (1) we show that for large $N$ it is given by the sine-kernel [see Eq. (\ref{sine_kernel})], 
which is well known in random matrix theory. In the two regions (2) and (3), we show that it takes different nontrivial scaling forms which we compute explicitly [see Eqs. (\ref{kernel_kf1}) and (\ref{Kp3_2}) respectively].

  \vspace*{0.5cm}
  
\subsubsection{Higher dimensions $d>1$} 
  
In this case, the Fermi surf is the product of two $d$-dimensional spheres defined by $|{\bf x}| = 1$ and $|{\bf p}| = k_F$ in position and momentum space respectively. Inside the Fermi surf, i.e. for $|{\bf x}| < 1$ and $|{\bf p}| < k_F$, the Wigner function is given, in the large $N$ limit, by the LDA prediction [see Eq. (\ref{LDA})], i.e.
  \bea \label{LDA_ddim}
  W_N({\bf x}, {\bf p}) \approx \frac{1}{(2 \pi)^d} \;.
  \eea
Here, we analyse the behavior of the Wigner function near the hard wall in space at a point close to ${\bf x}_w$ with $|{\bf x}_w| = 1$, i.e. the analogue of region III in the one-dimensional case (see Fig. \ref{fig:3dplots}). The result in the large $N$ limit is given by formula \eqref{radial4} where the geometry in momentum space is depicted in Fig. \ref{fig:2dnt}. Remarkably, the large $N$ scaling form of the Wigner function is independent of the spatial dimension. 
Note that a similar $d$-independence holds also for the Wigner function in the case of a smooth potential \cite{DDMS2018}, although in this case the Wigner function is  
given by a completely different formula.

\vspace*{0.5cm}   
   
The paper is organized as follows. In Section \ref{sec:wigner1d} we compute exactly the Wigner function for $N$ fermions in a one-dimensional hard box, which we then analyse in detail in the large $N$ limit. 
In Section~\ref{sec:mom1d} we focus on the statistics of momenta for $N$ fermions in a one-dimensional hard box and obtain explicit formulae for the density as well as the kernel in the large $N$ limit. In Section~\ref{sec:wignerd}, we compute the Wigner function for $N$ fermions in a $d$-dimensional hard-box, with a special focus on its large $N$ limiting form near the wall, while Section~\ref{sec:conclusion} contains our conclusions. Some further discussions have been left in five Appendices. In Appendix \ref{app:single}, we recall the semi-classical interpretation of the Wigner function for a single particle in a hard box, while Appendix \ref{app:WII} is devoted to the asymptotic analysis of the scaling function describing the Wigner function in region II. In Appendix \ref{app:single-hardwall} and \ref{app:single-hardwall_d} we present the exact computation of the Wigner function in the presence a single hard wall in $d=1$ (in Appendix \ref{app:single-hardwall_d}) and in higher dimensions (in Appendix \ref{app:single-hardwall_d}). Finally, in Appendix \ref{app:Bessel} we gave some details about the Wigner function for the $d$-dimensional spherical hard box.

              \section{Wigner function for fermions in a hard box in $d=1$}
    \label{sec:wigner1d}
    
 We start with the Wigner function $W_N(x,p)$ of $N$ noninteracting spinless fermions in a one-dimensional hard box (\ref{def_V_Box}) in their ground state. We first obtain an exact expression of $W_N(x,p)$ for finite $N$ in subsection \ref{sec:WN_initeN}, which we then analyse in the large $N$ limit in subsection \ref{sec:W_largeN}.

    \subsection{Exact results for finite $N$} \label{sec:WN_initeN}
    \subsubsection{Eigenstates}
    
The single particle Hamiltonian (\ref{def_H}) in a hard-box potential (\ref{def_V_Box}) reads, in $d=1$, setting $m=\hbar=1$ 
\bea \label{H_1d_box}
\hat H = -\frac{1}{2} \partial^2_x + V(x) \quad, \quad 
V(x) = 
\begin{cases}
&0 \quad, \quad -R \leq  x \leq R \;,\\
& + \infty \quad, \quad\quad\; |x| > R \;.
\end{cases}
\eea   
The single-particle eigenfunctions of $\hat H$, and associated eigen-energies, read in position representation 
\bea
  \phi_n(x) = \sqrt{\frac{1}{R}}\sin\left(\frac{n\pi}{2R}(x+R)\right){\mathbbm 1}_{[-R,R]}(x) \;, \;\quad \quad \epsilon_n= \frac{k_n^2}{2}=\frac{\pi^2}{8R^2} n^2 \;, \;  n\in\mathbb{N}^* \;,  \label{eq:wfposa} 
\eea
where $\mathbbm{1}_{[a,b]}(x)$ denotes the indicator function of the interval $[a,b]$. For later purposes, it is also useful to compute the eigenfunctions in the momentum representation where they are given by
\bea
 \hat \phi_n(p) = \frac{1}{\sqrt{2 \pi}}\int_{-R}^{R}  e^{-i p x} \phi_n(x) \,  dx= 4 \sqrt{\frac{R}{2 \pi}} \frac{n \pi}{[n^2 \pi^2 - 4 (pR)^2]} \sin{\left({pR} - n \frac{\pi}{2} \right)} e^{i(n+1)\frac{\pi}{2}} \;.
 \label{eq:wfmoma}
\eea
Note that there is no divergence at $pR = \pm n\pi/2$. Furthermore, one can check the normalization condition, $\int_{-\infty}^\infty|\hat \phi_n(p)|^2\, dp = 1$, using the identity
\bea \label{id}
8 n^2 \pi \int_{-\infty}^\infty \frac{1}{((n\pi)^2 - 4x^2)^2} \sin^2\left(x - n \frac{\pi}{2} \right)\,  dx = 1 \;, \;  n\in\mathbb{N}^* \;.
\eea

\subsubsection{Wigner function}

In the following, for simplicity and without loss of generality, we set $R=1$, which amounts to rescaling all the positions by $R$ and the momenta by $1/R$. In the ground-state, the $N$ lowest energy levels $\epsilon_n$ in (\ref{eq:wfposa}) are occupied, 
up to the Fermi energy $\mu = k_F^2/2$, where $k_F=N\pi/2$ is the Fermi wave vector. The $N$-body ground-state wave function $\Psi_0(x_1, \cdots, x_N)$ is given by the $N \times N$ Slater determinant built from the single particle eigen-state
\bea \label{Psi0_1d}
\Psi_0(x_1, \cdots, x_N) = \frac{1}{\sqrt{N!}} \det_{1 \leq k,\ell \leq N} \phi_k(x_\ell) \;.
\eea
Inserting this expression (\ref{Psi0_1d}) in the definition of the $N$ particle Wigner function $W_N(x,p)$ in Eq. (\ref{def_WN}), one can show that it can be written as (see e.g. \cite{DDMS2018})
  \begin{align}
    W_N(x,p) =& \frac{1}{2\pi}\sum_{n=1}^N \int_{-\infty}^\infty \phi^*_n(x+y/2)\phi_n(x-y/2) e^{ipy}\, dy\label{eq:Wa} \;.
  \end{align}
Note that because of the indicator function in the eigenfunction in (\ref{eq:wfposa}) with $R=1$, the support of the integral over $y$ in (\ref{eq:Wa}) is actually $-2+2|x| \leq y \leq 2 - 2|x|$. The generic term of this sum over $n$, which corresponds to the Wigner function of a single particle in the $n$-th excited state (\ref{eq:wfposa}) can be written as, using $\sin a \sin b = [\cos(a-b)-\cos(a+b)]/2$,  
\bea
  &&  \frac{1}{2\pi}\int_{-\infty}^\infty \phi^{*}_n(x+y/2)\phi_n(x-y/2) e^{ipy}\, dy = \frac{1}{4\pi}\int_{-2+2|x|}^{2-2|x|} e^{ipy}\left[\cos(n\pi y/2)-\cos(n\pi(x+1))\right]\, dy \nonumber \\
    &&=  \frac{1}{4\pi}\left[\int_{-2+2|x|}^{2-2|x|} e^{ipy}\cos(n\pi y/2)\, dy - 2\cos(n\pi(x+1))\frac{\sin(2p(1-|x|))}{p}\right] \:. \label{eq:psipsia} 
 \eea   
This form (\ref{eq:psipsia}) will be useful in the following to analyse the large $N$ limit of $W_N(x,p)$ in Eq. (\ref{eq:Wa}). Note that the remaining integral over $y$ in (\ref{eq:psipsia}) can be explicitly performed, yielding
\bea
&&   \frac{1}{2\pi}\int \phi^{*}_n(x+y/2)\phi_n(x-y/2) e^{ipy}\, dy \nonumber \\
&&= \frac{1}{4\pi}\Bigg(\frac{1}{p-n\pi/2} \sin\left((p-n\pi/2)(2-2|x|)\right) + \frac{1}{p+n\pi/2} \sin\left((p+n\pi/2)(2-2|x|)\right)  \nonumber	 \\
&&- 2\cos(n\pi(x+1))\frac{\sin(p(2-2|x|))}{p}\Bigg) \label{eq:psipsia2} 	\;.
\eea
As discussed in Ref. \cite{CKM1991,BDR2004} (see also Appendix \ref{app:single}), one can interpret the first two contributions in (\ref{eq:psipsia2}) in terms of a classical phase-space picture, while the last term in (\ref{eq:psipsia2}) comes from interferences and has a purely quantum mechanical origin.

Inserting Eq.~(\ref{eq:psipsia}) into Eq.~(\ref{eq:Wa}) and permuting the integral with the sum leads to
  \begin{align}
    \begin{split}   W_N(x,p) =& \frac{1}{4\pi} \int_{-2+2|x|}^{2-2|x|} e^{ipy} \, \left(\sum_{n=1}^N \cos({n\pi y/2})\right)\, dy - \left(\sum_{n=1}^N\cos(n\pi(x+1))\right)\frac{\sin(p(2-2|x|))}{2\pi p}\;.
      \end{split}\label{eq:Wsuma}
  \end{align}
Performing the sums over $n$ using the identity  
\bea \label{DN}
\sum_{n=1}^N \cos{n \pi z} =  \pi D_N(\pi z) - \frac{1}{2} \quad, \quad {\rm where} \quad  D_N(z)= \frac{\sin{\left((N+1/2)  z \right)}}{2\pi\sin{\left( \frac{ z}{2}\right)}}\, 
\eea 
is the Dirichlet kernel \cite{footnote1}, one obtains (note that the contributions due to the term $-1/2$ in the first identity in (\ref{DN}) cancel between the two sums over $n$ in Eq. (\ref{eq:Wsuma}))
    \begin{equation}
        {\begin{split} W_N(x,p) &=  \frac{1}{\pi} \int_{0}^{\pi(1-|x|)} \cos\left(\frac{2p}{\pi}u\right) D_{N}(u) \, du - D_{N}(\pi (1+x))\frac{\sin(2p(1-|x|))}{2 p}\,. \end{split} }\label{eq:Wua}
    \end{equation} 
An alternative expression for $W_N(x,p)$, which will also be useful in the following, can be obtained by performing first the integral over $y$ in Eq. (\ref{eq:Wsuma}) before the sum over $n$. This yields   
\bea \label{Wigner_alter}
W_N(x,p) = \frac{1}{2 \pi^2} \sum_{k=-N}^N \frac{\sin{\left( (k+\frac{2p}{\pi})\pi(1-|x|)\right)}}{k+\frac{2p}{\pi}} - D_{N}(\pi (1+x))\frac{\sin(2p(1-|x|))}{2 p}\ \;.
\eea

\begin{figure}[t]
\subfloat[$\rho_N(x)$]{%
 \includegraphics[width=0.4\textwidth]{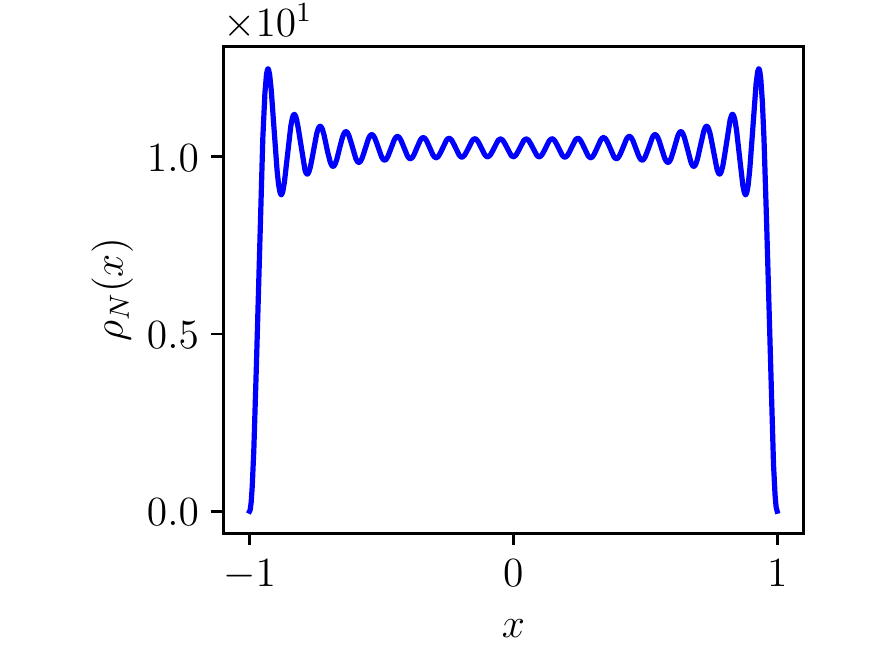}%
}~
\subfloat[$\hat \rho_N(p)$]{%
  \includegraphics[width=0.4\textwidth]{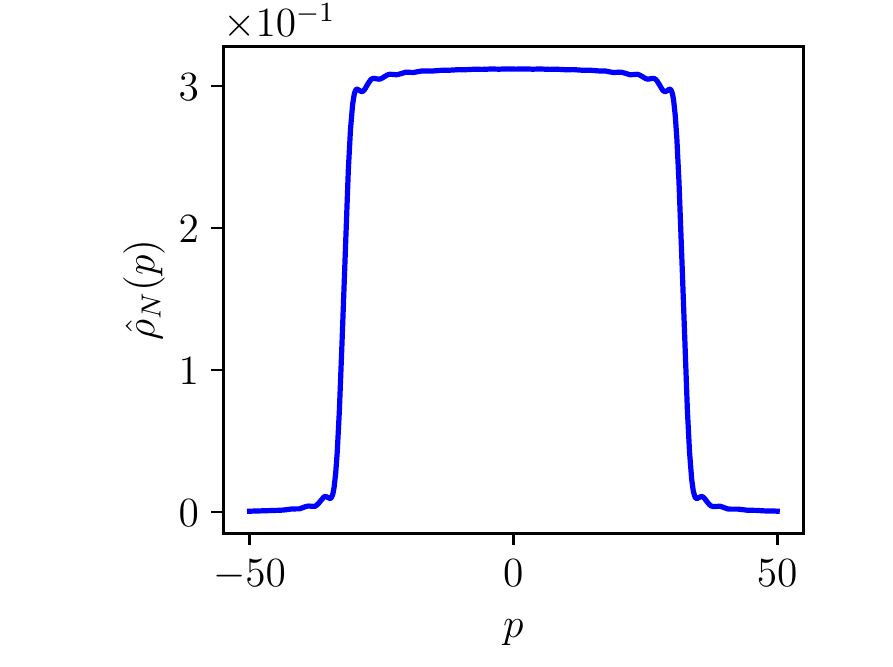}%
}\hfill
\caption{Density in position space (a) and in momentum space (b) for $N=20$ fermions.}\label{fig:density}
\end{figure}

A 3$d$-plot of the Wigner function $W_N(x,p)$ given in (\ref{eq:Wua}) is shown in Fig.~\ref{fig:3dplots} for $N=20$ fermions. This figure shows striking peaks close to $p=0$ and they were the main
subject of studies of the previous works on the Wigner function for fermions in the presence of hard-wall potentials \cite{ADSR1987,CKM1991} (see also Ref. \cite{berry1} for a discussion in a more general context). These Friedel-type oscillations \cite{ADSR1987} near $p=0$ arising from the second term in Eq. \eqref{eq:Wua} are further discussed in Appendix \ref{app:single}. Apart from these peaks, the Wigner function is roughly constant inside the rectangle delimited by the Fermi surf \eqref{Fermi_surf} (see also Fig. \ref{fig:fermisurf}), which is consistent with the LDA prediction (\ref{LDA}). It also shows non-trivial oscillating behaviors at the edge of the Fermi surf, which we will analyse below in detail in the large $N$ limit. It is somewhat easier to visualize these edge behaviors for the densities $\rho_N(x)$ and $\hat \rho_N(p)$. Indeed, by integrating $W_N(x,p)$ over $p$, and for this purpose it is convenient to use the expression in (\ref{eq:Wsuma}), one obtains
\bea \label{density_x}
\rho_N(x) = \frac{2N+1}{4} - (-1)^N \frac{\cos{\left( (N+1/2) \pi x \right)}}{4\cos{\left( \frac{\pi x}{2}\right)}} \;,
\eea
recovering the result of \cite{lacroix2018non}. On the other hand, by integrating over $x$ one finds
\bea \label{density_p}
\hat \rho_N(p) = \sum_{k=1}^N\frac{4 \pi  k^2 \left((-1)^{k+1} \cos (2 p)+1\right)}{\left(\pi ^2 k^2-4 p^2\right)^2}\,.
\eea
It is interesting to note that, for $N$ finite and large $p \gg N$, the momentum density has an $1/p^4$ algebraic tail  
\bea \label{tail_rhoN}
\hat \rho_N(p) = \frac{\pi}{96 p^4} N (N+1)(2N+1-3(-1)^N \cos{2p}) + O (1/p^6) \;,
\eea
which, neglecting the oscillating term, gives the formula (\ref{tail_rhoN_intro}) given in the introduction. As discussed earlier, this tail also appears in quantum particle systems with contact repulsion. Here, for noninteracting fermions in the presence of an infinite wall, the eigenfunctions vanish near the wall as $|y| \theta(y)$ where $y$ denotes the distance from the wall [see Eq. (\ref{eq:wfposa})]. Hence, in Fourier space, they behave as $1/p^2$ at large $p$ [see Eq. (\ref{eq:wfmoma})], leading to the $1/p^4$ tail in the momentum density. In Fig.~\ref{fig:density} a) and b) we show a plot of $\rho_N(x)$ in (\ref{density_x}) and $\hat \rho_N(p)$ in (\ref{density_p}) for $N=20$ fermions. In both cases, the densities are uniform over a finite support, displaying oscillations which are enhanced close to the edges. Note also that the shape of these oscillations, in the position and the momentum space, are seemingly rather different, which will be confirmed by our computations below.

%

\subsection{Asymptotic results for large $N$}\label{sec:W_largeN}

We now analyse the Wigner function $W_N(x,p)$ given in (\ref{eq:Wua}) in the large $N$ limit and analyse separately the four regions I, II, III and IV discussed above (see Fig. \ref{fig:fermisurf}) in four different subsections. Without loss of generality, we will restrict the analysis to the first quadrant of the $x$-$p$ plane as the Wigner function is symmetric both with respect to $x$ and $p$.

\subsubsection{Region I:  bulk $(-1<x<1$ and $-k_F< p < k_F)$}

We first consider the bulk region I, i.e. $-1< x < 1$ and $-k_F <p < k_F$, with $k_F = N \pi/2$. We thus set $p = \tilde p \, k_F = \tilde p N\pi/2$ and study $W_N(x,p=\tilde p N\pi/2)$ for large $N$. It is easy to see that the second term in Eq. (\ref{eq:Wua}) goes to zero, while the first one, as we will see, gives a finite contribution. Hence one has
\bea \label{WN_I1}
W_N\left(x, p = \tilde p \, \frac{N \pi}{2}\right) \simeq \frac{1}{4 \pi} \int_0^{2(1-|x|)} dy\, \cos{\left( \tilde p\,y\,N\frac{\pi}{2}\right)} \frac{\sin{\left((N+\frac{1}{2}) \frac{\pi y}{2} \right)}}{\sin{\left(\frac{\pi y}{4} \right)}} \;.
\eea
Performing the change of variable $y = 2v/(\pi N)$ we get 
\bea \label{WN_I2}
W_N\left(x, p = \tilde p \frac{N \pi}{2}\right) \simeq \frac{1}{\pi^2} \int_0^\infty \frac{dv}{v} \, \cos(\tilde p\, v) \sin (v) = \frac{1}{2 \pi} \Theta(1-|\tilde p|) \;.
\eea
Note that the integral over $v$ in (\ref{WN_I2}) has been simply evaluated by writing $\cos(\tilde p v) \sin (v) = (\sin((\tilde p+1)v) - \sin((\tilde p-1)v))/2$ and using $\int_0^\infty \sin(av)/v\, dv = \frac{\pi}{2} \sgn(a)$. This result (\ref{WN_I2}) leads to the behavior announced in Eq. (\ref{region_I}), which coincides with the prediction of the LDA (\ref{LDA}). 

\begin{figure}[t]
      \includegraphics[width=0.4\textwidth]{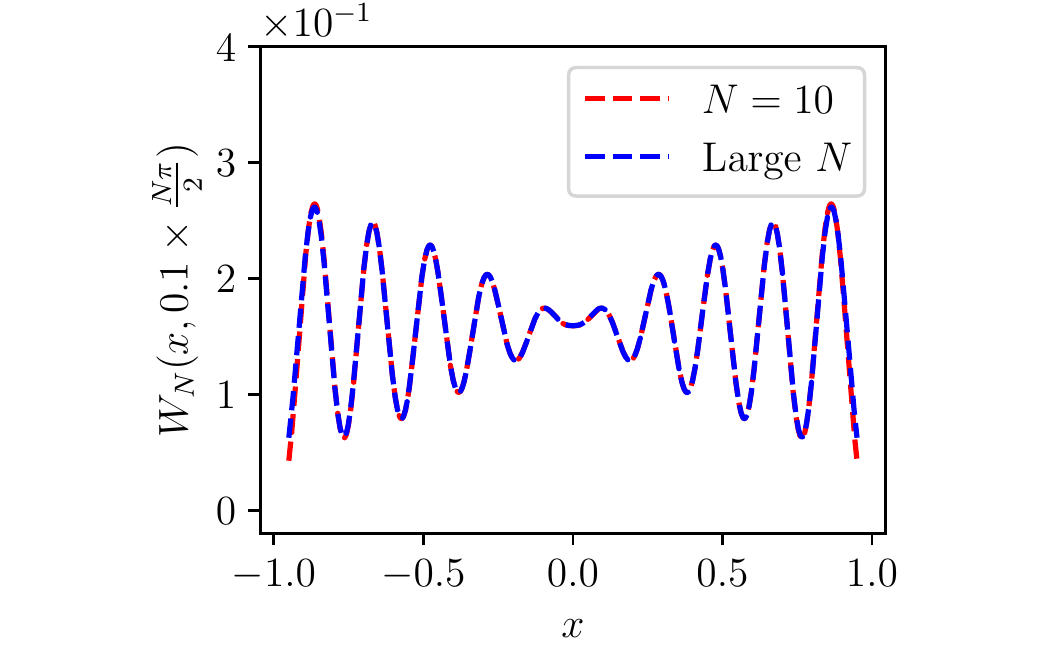}
      \caption{Plot of the Wigner function $W_N\left(x,p\right)$ vs $-1<x<1$ (i.e. in the bulk of the box) for fixed  $p=0.1\times\frac{N\,\pi}{2}$ (see region I in figure \ref{fig:fermisurf} and \ref{fig:3dplots}). The red dashed  line corresponds to the exact value of $W_N(x,p)$ for $N=10$ while the blue dashed line corresponds to the formula (\ref{bulk_1overN}) which includes the first $1/N$ corrections to the LDA prediction $W_N(x,p) \approx 1/(2 \pi)$. As $N$ increases, the Wigner function approaches this constant value $1/(2\pi)$, while exhibiting oscillations which, as we see, are accurately described by the $1/N$ corrections in (\ref{bulk_1overN}). }
      \label{fig:bulk}
    \end{figure}

With a bit more work, it is possible to obtain the $1/N$ correction to this constant value $1/(2 \pi)$ (\ref{WN_I2}) in the bulk. It reads, up to terms of order $O(1/N)$
\bea \label{bulk_1overN}
&&W_N\left(x, p =\tilde p \frac{N \pi}{2}\right) \simeq \frac{1}{2 \pi} \Theta(1-|\tilde p|) \\
&+& \frac{1}{N 4 \pi^2} \frac{1}{\cos{\left(\frac{\pi x}{2}\right)}} \frac{1}{\tilde p(\tilde p^2-1)} \left[(1+\tilde p)\cos{\left(\frac{\pi}{2}(1-|x|)(2N(\tilde p-1)-1)\right)} + (\tilde p-1)\cos{\left(\frac{\pi}{2}(1-|x|)(2N(\tilde p+1)+1)\right)}  \right] \;.  \nonumber
\eea
We have checked numerically that this formula (\ref{bulk_1overN}) provides a very good approximation of the exact Wigner function (\ref{eq:Wua}) or (\ref{Wigner_alter}) for all values of $x$ and $p$, provided $x$ is
not too close to the hard wall, for $N \gtrsim 10$ (see Fig.~\ref{fig:bulk}). 

\subsubsection{Region II:  momentum edge ($-1<x<1$ and $p = k_F + O(1)$)}

In region II, $x$ is in the bulk, $-1<x<1$ (i.e., far from the wall) but $p$ is close to $k_F = N \pi/2$ and we thus set $p = k_F + q \pi/2 = (\pi/2)(N + q)$, with $q = O(1)$. In this regime, and in the limit of
large $N$, the second term in (\ref{eq:Wua}) vanishes as $N \to \infty$ while, the first term remains finite in the limit $N \to \infty$. Hence the Wigner function in (\ref{eq:Wua}) reads in regime II
\bea \label{WN_II1}
W_N\left(x, p = \frac{\pi}{2}(N+q)\right) \simeq \frac{1}{\pi} \int_{0}^{\pi (1-|x|)} \cos\left((N+q)u\right) D_N(u)\,  du\;.
\eea
Using the explicit expression of the Dirichlet kernel $D_N(u)$ from Eq. (\ref{DN}) together with the trigonometric identity $2\cos(a)\sin(b)=\sin(a+b)-\sin(a-b)$, the expression in (\ref{WN_II1}) becomes
\bea  \label{WN_II2}
W_N\left(x, p = \frac{\pi}{2}(N+q)\right) \simeq \frac{1}{2\pi} \int_{0}^{\pi (1-|x|)}  \left[D_{2N+q}(u) -  D_{q-1}(u)\right] \; du \;.
\eea
Note that the Dirichlet kernel $D_N(z)$ in (\ref{DN}), while originally defined for integer values $N$, can be straightforwardly analytically 
continued to real values of $N$ -- see the second equality in (\ref{DN}). Besides, in the limit $N \to \infty$ one can easily show that 
\bea \label{eq:diricheletDelta}
\lim_{N \to \infty} \int_0^a  D_N(x) f(x)\, dx = \frac{f(0)}{2} \;,
\eea
for any smooth function $f(x)$ \cite{footnote2}. Using this identity (\ref{eq:diricheletDelta}), we see that the expression in (\ref{WN_II2}) has a good large $N$ limit, namely
\bea \label{eq:EII}
\lim_{N \to \infty} W_N\left(x, p = \frac{\pi}{2}(N+q)\right)  =  \frac{1}{2 \pi} {\cal W}_{\rm II}(x,q) \quad, \quad {\cal W}_{\rm II}(x,q) = \frac{1}{2} -  \int_{0}^{\pi (1-|x|)}  D_{q-1}(u)\,du \;.
\eea

\begin{figure}[t]
      \includegraphics[width=0.4\textwidth]{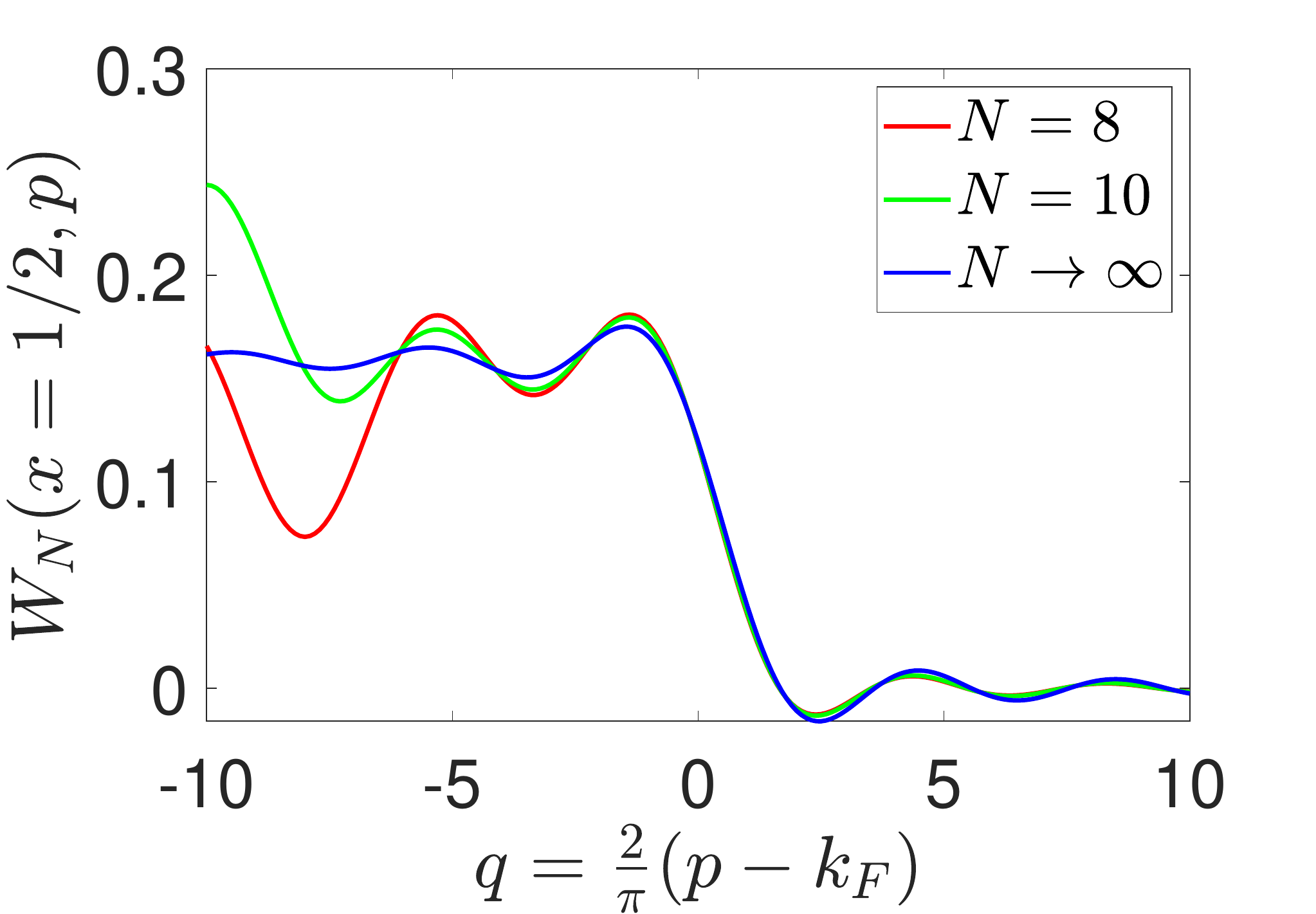}
      \caption{Exact and large $N$ description of a slice of the Wigner function $W_N\left(x,p\right)$ at the momentum edge $p=k_F+\frac{\pi}{2}\,q$ with $q=O(1)$ along the line $x=1/2$ (see region II in figure \ref{fig:fermisurf} and \ref{fig:3dplots}). As $N$ increases, the Wigner function approaches the scaling function ${\cal W}_{\rm II}(x,q)$ given by (\ref{region_II}) (blue line). }
      \label{fig:edgemom}
    \end{figure}

An alternative expression for the Wigner function in this regime, and thus of the scaling function ${\cal W}_{\rm II}(x,q)$, can be obtained by starting from the expression for $W_N(x,p)$ given in Eq. (\ref{Wigner_alter}), where, we recall
that in this regime II, the last term can be neglected compared to the sum over $k$. Setting $p = (\pi/2)(N+q)$, performing the change of variable $m = k +N$ in the sum and taking the limit
$N \to \infty$ one finds 
 \begin{eqnarray}\label{eq:EII_alter}
 \lim_{N \to \infty}W_N(x,p = \frac{\pi}{2}(N + q)) = \frac{1}{2 \pi} {\cal W}_{\rm II}(x,q) \quad, \quad {\cal W}_{\rm II}(x,q) = \frac{1}{\pi} \sum_{m=0}^\infty \frac{\sin((m+q) \pi (1-|x|))}{m+q}\,,
 \end{eqnarray}  
 as announced in the introduction \eqref{region_II}. A plot of the scaling function ${\cal W}_{\rm II}(x,q)$ is shown in Figs. \ref{fig:3dplots} b) and \ref{fig:edgemom}.

Although the two formulae (\ref{eq:EII}) and (\ref{eq:EII_alter}) may look different, one can check that they indeed coincide. It is interesting to analyse the large $|q|$ behavior of this scaling function ${\cal W}_{\rm II}(x,q)$. As shown in Appendix \ref{app:WII}, this is conveniently done starting from the expression (\ref{eq:EII}) and we get
\bea \label{largek_II}
{\cal W}_{\rm II}(x,q) = \Theta(-q) + \frac{1}{q} \frac{\sin{\left(\pi q(1-|x|) + \frac{\pi}{2}|x| \right)}}{2 \pi \cos{\frac{\pi x}{2}}} + O(1/q^2) \quad, \quad |q| \to + \infty \;. 
\eea
%
Note that on both sides, i.e. for $q \to - \infty$ and $q \to +\infty$, the Wigner function shows oscillations (around its constant value) whose amplitude decays quite slowly, i.e. $\sim 1/q$, as $q \gg 1$. In addition, we see from (\ref{largek_II}) that the Wigner function, namely for $q \to +\infty$, can actually be negative in this region II. These features are in marked contrast with the behavior found for smooth potentials where the Wigner function is described by Eq. \eqref{scal2}. Indeed, in this case the decay is typically faster than exponential and the Wigner function remains positive. 

Finally, although our analysis in this regime holds for $-1<x<1$, i.e. sufficiently far from the wall, it is interesting to study the limiting behavior of ${\cal W}_{\rm II}(x,q)$ as $x \to 1$, which amounts to study the Wigner function near the top right corner of the Fermi surf in Fig. \ref{fig:fermisurf}. Indeed, from the representation in Eq. \eqref{eq:EII} one immediately obtains that 
\bea \label{top_right}
{\cal W}_{\rm II}(x,q) \sim \frac{1}{2} \quad, \quad x \to 1 \;,
\eea
which is thus half the value of the Wigner function in the bulk [see Eq. (\ref{region_I})].



\subsubsection{Region III: near the wall ($1-x = O(1/k_F)$ and $p = O(k_F)$)}

\begin{figure}[t]
      \centering
      \includegraphics[width=0.4\textwidth]{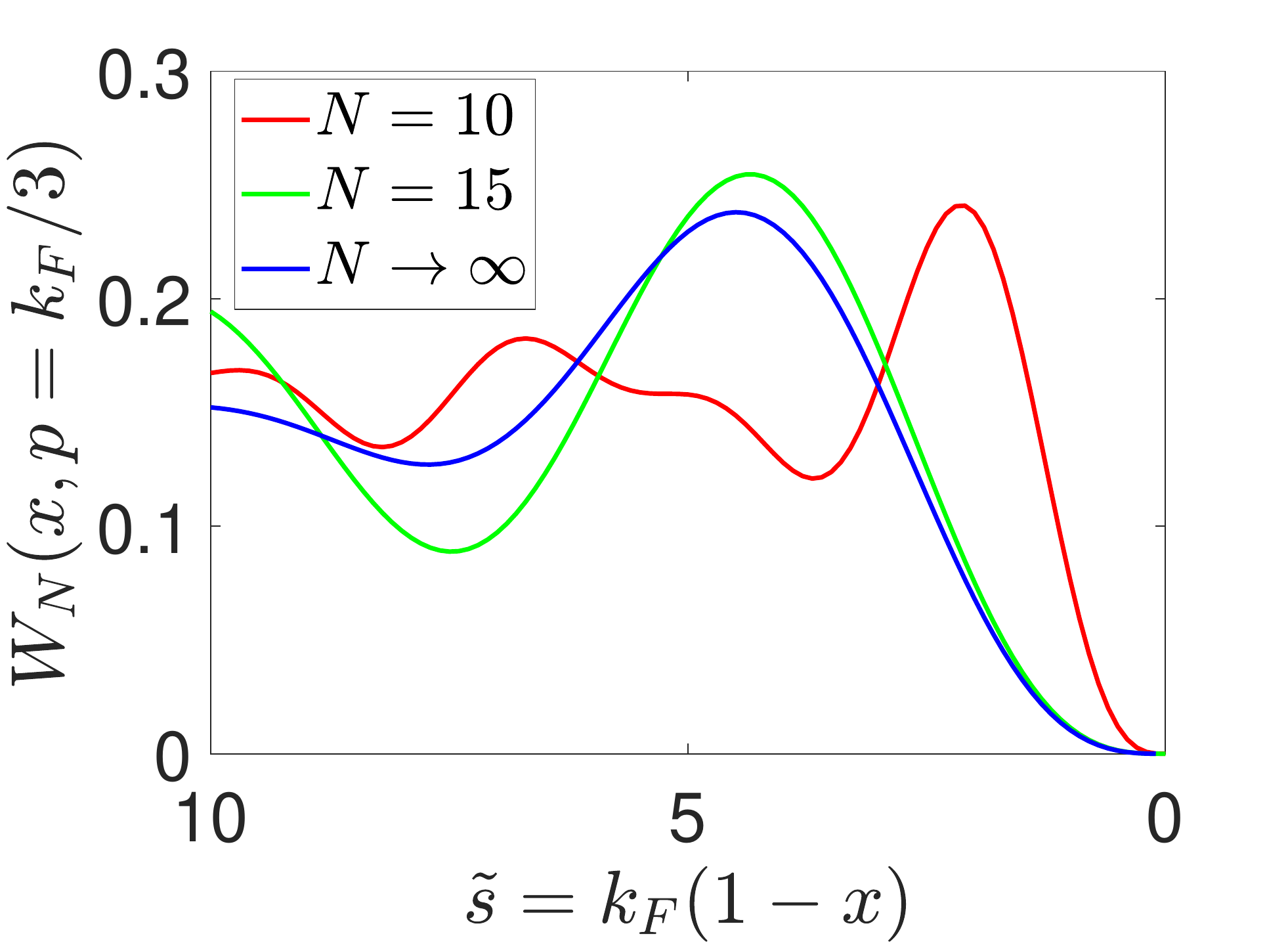}
      \caption{Exact and large $N$ description of a slice of the Wigner function $W_N(x,p)$ near the wall $x=1-\frac{\tilde s}{k_F}$ with $\tilde s=O(1)$ along the line $p=\tilde p\,k_F$ with $\tilde p=\frac{1}{3}$ (see region III in figure \ref{fig:fermisurf} and \ref{fig:3dplots}). As $N$ increases, the Wigner function approaches the scaling function ${\cal W}_{\rm III}(\tilde s,\tilde p)$ given by (\ref{region_III}) (blue line).}
      \label{fig:wall}
    \end{figure}

We now analyse the Wigner function in region III, i.e. close to the hard wall (see Fig.~\ref{fig:fermisurf}). In this regime, it is convenient to start from Eq. (\ref{eq:Wua}) and set 
$x = 1 - \tilde s/k_F$, with $\tilde s>0$, and $p = \tilde p \, k_F$, with $k_F = N \pi/2$, to obtain
  \begin{align}
  W_N(x = 1 -  \tilde s/k_F, p = \tilde p \, k_F)  =
   \frac{1}{\pi} \int_{0}^{2\tilde s/N} \cos\left(N\tilde p u\right) D_N(u)\,  du - D_N\left(\frac{2\tilde s}{N}\right)\frac{\sin(2\tilde p \tilde s)}{\pi N\tilde p}\,. \label{EIII}
  \end{align}
Performing the change of variable $u \to u/N$ and using $D_N(x/N)\sim N \sin(x)/(\pi x)$ as $N \to \infty$, one obtains straightforwardly from Eq. (\ref{EIII}) that $W_N(x = 1 -  \tilde s/k_F, p = \tilde p \, k_F)$ reads, in the limit $N \to \infty$, keeping $\tilde s$ and $\tilde p$ fixed 
\bea  \label{EIII_2}
\lim_{N \to \infty}  W_N(x = 1 -  \tilde s/k_F, p = \tilde p \, k_F) = \frac{1}{2 \pi} {\cal W}_{\rm III}(\tilde s, \tilde p) \;,
\eea
with the scaling function  ${\cal W}_{\rm III}(\tilde s, \tilde p)$ given in Eq. (\ref{region_III}). A plot of this function is shown in Figs. \ref{fig:3dplots} c) and \ref{fig:wall}. 

It is interesting to study the asymptotic behaviors of this scaling function ${\cal W}_{\rm III}(\tilde s, \tilde p)$ in various limits. Let us first consider the small $\tilde s$ behavior, i.e. very near the wall. In this limit, it is easy to obtain from the expression given in Eq. (\ref{region_III}) that
  \begin{align}\label{W3_smalls}
      \mathcal{W}_{\text{III}}(\tilde s,\tilde p) = \frac{16}{9\pi}\, \tilde s^3 + O(\tilde s^5) ,\quad \tilde s\rightarrow 0 \;,
    \end{align}
independently of $\tilde p$. The large $\tilde s$ behavior, i.e. in a region towards the bulk, of ${\cal W}_{\rm III}(\tilde s, \tilde p)$ is a bit more subtle. Indeed, focusing on the case $\tilde p >0$, we see on Eq. (\ref{region_III}) that $W_{\rm III}(\tilde s, \tilde p)$ exhibits different behaviors depending on $\tilde p > 1$ or $\tilde p < 1$, since the Sine integral function ${\rm Si}(x)$, being an odd function, behaves differently for $x \to + \infty$ and $x \to -\infty$. Namely, one has
\bea \label{Si_asympt}
{\rm Si}(x) =\sgn{(x)}\,\frac{\pi}{2} - \frac{\cos x}{x} + O(1/x^2) \;, \; x \to \pm \infty \;.
\eea 
Hence one finds
\bea \label{W3_larges}
{\cal W}_{\rm III}(\tilde s, \tilde p) = 
\begin{cases}
&\dfrac{1}{\tilde s}\,\dfrac{(\tilde p+1) \cos (2 (\tilde p-1) \tilde s)+(\tilde p-1) \cos (2 (\tilde p+1) \tilde s)}{2 \pi  \tilde p \left(\tilde p^2-1\right) }+ O(1/\tilde s^2) \quad, \quad \quad\;\; \tilde s\to \infty \quad {\rm for } \quad \tilde p > 1 \;, \\
& \\
&1+\dfrac{1}{\tilde s}\,\dfrac{(\tilde p+1) \cos (2 (\tilde p-1) \tilde s)+(\tilde p-1) \cos (2 (\tilde p+1) \tilde s)}{2 \pi  \tilde p \left(\tilde p^2-1\right) }+ O(1/\tilde s^2) \quad, \quad \tilde s\to \infty \quad {\rm for } \quad 0<\tilde p < 1 \;.
\end{cases}
\eea
Such different behaviors for $\tilde p < 1$ (i.e., $p<k_F$) and $\tilde p > 1$ (i.e., $p>k_F$) as $s \to \infty$, i.e. far from the wall, are of course expected given the behavior of the Wigner function in the bulk, i.e. in the region I [see Eq. (\ref{region_I}) and Fig. \ref{fig:fermisurf}]. For $\tilde p \simeq 1$, there is an interesting crossover region, which is discussed below.  

In the limit of vanishing momentum $\tilde p = 0$, the Wigner function in this regime takes the simple form
\bea \label{W3p0}
{\cal W}_{\rm III}(\tilde s, \tilde p =0)= \frac{2}{\pi} \left( {\rm Si}(2\tilde s) - \sin{2\tilde s}\right) \;.
\eea
In particular, in the limit of large $\tilde s$ one has
\bea \label{W3p0smalls}
{\cal W}_{\rm III}(\tilde s, \tilde p =0) = 1 - \frac{2}{\pi} \sin{(2\tilde s)} + O(1/\tilde s) \quad, \quad \tilde s \to +\infty  \;,
\eea 
which shows that, in this case, the spatial oscillations are not damped -- contrarily to the case $0<\tilde p<1$ (see the second line in Eq. (\ref{W3_larges})) where the oscillating term is multiplied by $1/\tilde s$ and thus decays as $\tilde s \to +\infty$. Finally, it is also interesting to study the behaviour of ${\cal W}_{\rm III}(\tilde s, \tilde p)$ for large $\tilde p$. From the explicit expression in Eq. (\ref{region_III}), and using the asymptotic behavior in Eq. (\ref{Si_asympt}), it is straightforward to obtain
 \begin{align}\label{W3_largep}
      \mathcal{W}_{\text{III}}(\tilde s,\tilde p) = \frac{1}{\tilde p^2} \,\frac{\cos (2 \tilde p \tilde s) (2 \tilde s \cos
   (2 \tilde s)-\sin (2 \tilde s))}{2 \pi  \tilde s^2} + O(1/\tilde p^3)\, ,\quad \tilde p \to +\infty \;,
    \end{align}
which, again, decays algebraically with $\tilde p$ (modulated by a periodic function), i.e., much slower than the faster than exponential decay found for smooth potentials [see Eq. (\ref{asymp_plus})]. However, this $1/\tilde p^2$ behavior is integrable, as it should since the total integral of $W_N(x,p)$ over $p$ yields the spatial density [see the first relation in Eq. (\ref{marginals})]. In fact, this implies, using the scaling for the spatial density near the wall in Eq. (\ref{rhox_2}), that ${\cal W}_{\rm III}(\tilde s, \tilde p)$ obeys the relation
\be  \label{constraint_WIII}
\int_{-\infty}^\infty \mathcal{W}_{\text{III}}(\tilde s,\tilde p) d\tilde p  = \frac{4}{N} \rho_N(x) \simeq 2\, F_1(\tilde s) \quad , \quad F_1(\tilde s)= 1 - \frac{\sin (2 \tilde s)}{2 \tilde s} \;.
\ee 
We have checked, using the explicit expressions for $ \mathcal{W}_{\text{III}}(\tilde s,\tilde p)$ in Eq. (\ref{region_III}) and $F_1(\tilde s)$ in Eq. (\ref{rhox_2}) that this identity (\ref{constraint_WIII}) is indeed satisfied.

\vspace*{0.5cm}

It turns out that the scaling function ${\mathcal{W}}_{\rm III}(\tilde  s, \tilde p)$ can be obtained directly by using the result for the limiting form of the kernel near the wall at $x=1$, obtained in Ref. \cite{lacroix2018non}. Indeed, in general, $W_N(x,p)$ can be written in terms of the kernel as \cite{DDMS2018} as given in \eqref{W_vs_Kernel}.
For the present box potential (\ref{def_V_Box}) in $d=1$ with $R=1$, setting $x = 1 - \tilde  s/k_F$ and $p = \tilde p k_F$ and performing the change of variable $z = k_F y$ in (\ref{W_vs_Kernel}) one has
\bea \label{W_vs_Kernel_2}
W_N(x=1-\tilde  s/k_F, p = \tilde p k_F) = \frac{1}{2 \pi} \int_{-2 \tilde  s}^{2 \tilde  s} dz \, e^{i \tilde p z} \frac{1}{k_F} K_N\left(1 - \frac{1}{k_F}(\tilde  s+\frac{z}{2}), 1 - \frac{1}{k_F}(\tilde  s-\frac{z}{2}) \right) \;.
\eea
In the limit of large $N$, one can then use that the kernel near the wall in (\ref{W_vs_Kernel_2}) takes the scaling form (see Ref. \cite{lacroix2018non}) of a ``reflected'' sine kernel (see also \cite{CMV2011})
\bea \label{limiting_kernel}
\frac{1}{k_F} K_N\left(1 - \frac{1}{k_F}(\tilde  s+\frac{z}{2}), 1 - \frac{1}{k_F}(\tilde  s-\frac{z}{2})  \right) \underset{N \to \infty}{\longrightarrow} \frac{1}{\pi} \left(\frac{\sin z}{z}-\frac{\sin (2\tilde  s)}{2\tilde  s} \right) \;.
\eea
Inserting this scaling form (\ref{limiting_kernel}) into (\ref{W_vs_Kernel_2}) and performing the integral over $z$ one obtains immediately
\bea \label{W_vs_Kernel_3}
W_N(x=1-\tilde  s/k_F, p = \tilde p k_F)  \underset{N \to \infty}{\longrightarrow} \frac{1}{2\pi} \, {\cal W}_{\rm III}(\tilde  s, \tilde p) \;,
\eea
where ${\cal W}_{\rm III}(\tilde  s, \tilde p)$ is given in (\ref{region_III}). This provides an alternative derivation of this scaling function -- which can be extended to higher dimensions $d>1$ as we will see later in Section \ref{sec:wignerd}. 

\vspace*{0.5cm}
We end this section by mentioning that the Wigner function ${\cal W}_{\rm III}(\tilde  s, \tilde p)$ in this regime III turns out to coincide exactly with the Wigner function for noninteracting
fermions on the half-line $x \geq 0$ in the presence of a hard-wall potential at $x=0$ (see Appendix \ref{app:single-hardwall-flat} for details). Furthermore, in Appendix \ref{app:single-hardwall-inv}, we
show how this Wigner function gets modified in the presence of an inverse-square repulsive potential near the origin.

 \subsubsection{Region IV: corner ($x \to 1$, $p - k_F \to \infty$ with the product $(1-x)(p-k_F)$ fixed)}

\begin{figure}[t]
      \centering
      \includegraphics[width=0.5\textwidth]{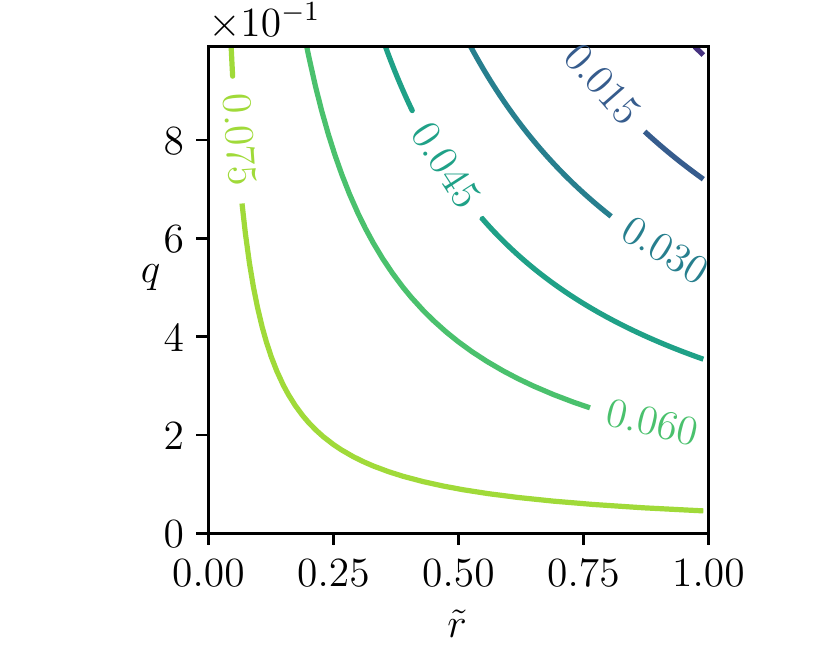}
      \caption{Contour plot of the large $N$ limit of the Wigner function $W_N\left(x,p \right)$ in the corner of the Fermi surf $k_F^{-1} \ll 1-x \ll 1$ and $1\ll |p - k_F| \ll k_F$ but keeping the product $(1-x)(p-k_F) = z$ fixed (see the region IV in figure \ref{fig:fermisurf} and \ref{fig:3dplots}). For illustrative purposes, we used the rescaled coordinates $x=1-\frac{\tilde r}{k_F^\alpha}$ and $p=k_F +q k_F^\alpha $ with $\alpha<1$. The hyperbolic curves $(1-x)(p-k_F) = \tilde r \,q= z$ appear clearly in the contour plot. The scaling function in the corner region  is given in (\ref{region_IV}).}
      \label{fig:corner3d}
    \end{figure}

In this regime, $x$ and $p$ are close to the top right corner (see Fig. \ref{fig:fermisurf}), we consider the ``mesoscopic'' scaling limit where $k_F^{-1} \ll (1-x) \ll 1$ and $1 \ll |k_F-p| \ll k_F$ but keeping the 
$(1-x)(p-k_F) = z$ fixed. To study this scaling region, it is is useful to set $1 - x = \frac{\tilde r}{k_F^\alpha}$ and $q=k_F +q k_F^\alpha$, with $0 < \alpha < 1$ and we recall that $k_ F= N \pi/2 \gg 1$.  
In this limit, it is more convenient to start from the expression for $W_N(x,p)$ given in Eq. (\ref{eq:Wua}). In this limit, again, it is easy to see that the second term in (\ref{eq:Wua}) is subdominant compared to the first one, i.e., 
   \begin{align} \label{corner1}
   W_N\left(x=1-\frac{\tilde r}{k_F^\alpha} ,p=k_F +q k_F^\alpha  \right)  &\simeq  \frac{1}{\pi} \int_{0}^{\frac{2^\alpha}{\pi^{\alpha-1}}\frac{\tilde r}{N^{\alpha}}} \cos\left(\left(N+\frac{ \pi^{\alpha-1}}{2^{\alpha-1}}N^\alpha q\right)u\right) D_N(u)\,  du\,.
  \end{align}
   Using the trigonometric identity $2\cos(a)\sin(b)=\sin(a+b)-\sin(a-b)$, we find that (\ref{corner1}) can be written as
   \begin{align}
         W_N\left(x=1-\frac{\tilde r}{k_F^\alpha} ,p=k_F +q k_F^\alpha  \right)  &\simeq  \frac{1}{2\pi} \int_{0}^{\frac{2^\alpha}{\pi^{\alpha-1}}\frac{\tilde r}{N^{\alpha}}}  D_{2N+\frac{ \pi^{\alpha-1}}{2^{\alpha-1}}q N^\alpha}(u) -  D_{\frac{ \pi^{\alpha-1}}{2^{\alpha-1}}q N^\alpha-1}(u)\, du\, .
   \end{align}
 Finally, performing the change of variable $v = u N^\alpha\frac{2^{\alpha-1}}{ \pi^{\alpha-1}}$ and taking the limit $N\rightarrow \infty$, we find 
\bea \label{corner2}
\lim_{N \to \infty}  W_N\left(x=1-\frac{\tilde r}{k_F^\alpha} ,p=k_F +q k_F^\alpha  \right) = \frac{1}{2\pi} {\cal W}_{\rm IV}(\tilde r\, q) \;,
 \eea 
 where the scaling function ${\cal W}_{\rm IV}(z)$ is given in Eq. (\ref{region_IV}). Note that this scaling function is independent of $\alpha$ in the range $0< \alpha < 1$. Its asymptotic behaviors are given by
 \bea \label{asympt_WIV}
{\cal W}_{\rm IV}(z) \sim
\begin{cases} \label{W4asympt}
&1 + \dfrac{1}{2 \pi z} \cos{(2z)} + O(1/z^2) \quad, \quad z \to - \infty \;, \\
& \\
&\dfrac{1}{2 \pi z} \cos(2z)  + O(1/z^2) \quad\quad \quad, \quad z \to +\infty \;.
\end{cases} 
 \eea
A plot of ${\cal W}_{\rm IV}(z)$ is shown in Fig.~\ref{fig:corner3d}.
   
By looking at Fig. \ref{fig:fermisurf} together with the scaling forms in Eqs. (\ref{region_II}) and (\ref{region_III}), we see that this region IV can be reached (i) either coming from regime II by letting $(1-x) \to 0$ and $q = (2/\pi)(p-k_F) \to \infty$, keeping $q (1-x) = \frac{2}{\pi}\,z$ fixed (where we recall that $z = (1-x)(p-k_F)$), or (ii) coming from the regime III by letting $\tilde s =(1-x)k_F \to \infty$ and $\tilde p = p/k_F \to 1$ with $\tilde s (\tilde p-1) = z$ fixed. In the first case (i), it is convenient to start from the expression for ${\cal W}_{\rm II}(x,q)$ given in Eq. (\ref{region_II}). Indeed, in this case, as $x \to 1$, the discrete sum over $m$ can be replaced by an integral  and one obtains
\bea \label{2to4}
{\cal W}_{\rm II}(x,q) \simeq \int_0^\infty \frac{\sin{\left(m \pi (1-x) + \pi q(1-x)\right)}}{m + q} \; dm = \frac{1}{2}  - {\rm Si}(2(1-x)q) = \frac{1}{2} - {\rm Si}(2(1-x)(p-k_F)) \;,
\eea 
which matches perfectly with the expression for ${\cal W}_{\rm IV}(z = (1-x)(p-k_F))$ in (\ref{region_IV}). Similarly, in the second case (ii), one immediately sees in the expression of ${\cal W}_{\rm III}(\tilde s, \tilde p)$ in Eq. (\ref{region_III}) that in the limit $\tilde s =(1-x)k_F \to \infty$ and $\tilde p = p/k_F \to 1$ with $\tilde s (\tilde p-1) = z$, the last term is subleading compared to the first two ones, which eventually gives
\bea \label{3to4}
{\cal W}_{\rm III}(\tilde s, \tilde p) \simeq \frac{1}{2} - {\rm Si}(2 \tilde s (\tilde p - 1)) =  \frac{1}{2} - {\rm Si}(2(1-x)(p-k_F))\;,
\eea
where we have used the first term of the asymptotic behavior of ${\rm Si}(x)$ given in (\ref{Si_asympt}) together with the fact that ${\rm Si}(-z) = - {\rm Si}(z)$. Therefore this expression (\ref{3to4}) also matches 
with the expression for ${\cal W}_{\rm IV}(z)$ in (\ref{region_IV}). Hence we see that this regime IV connects smoothly the regime II and the regime III (see Fig. \ref{fig:fermisurf}).

 \section{Momentum statistics for fermions in a hard box in $d=1$}
 \label{sec:mom1d}

In this section, we focus on the statistics of momenta for noninteracting fermions in a one-dimensional hard box. We first present the density and then the kernel in momentum space.

 \subsection{Density in momentum space}
 
We start with the exact expression for the density in momentum space, given in Eq. (\ref{density_p}), which we analyse in the large $N$ limit. We identify three different regimes which we analyse separately:  
 

\begin{figure}[htb]
\subfloat[Momentum density $\hat \rho_N(p)$ (\ref{density_p}).]{%
 \includegraphics[width=0.45\textwidth]{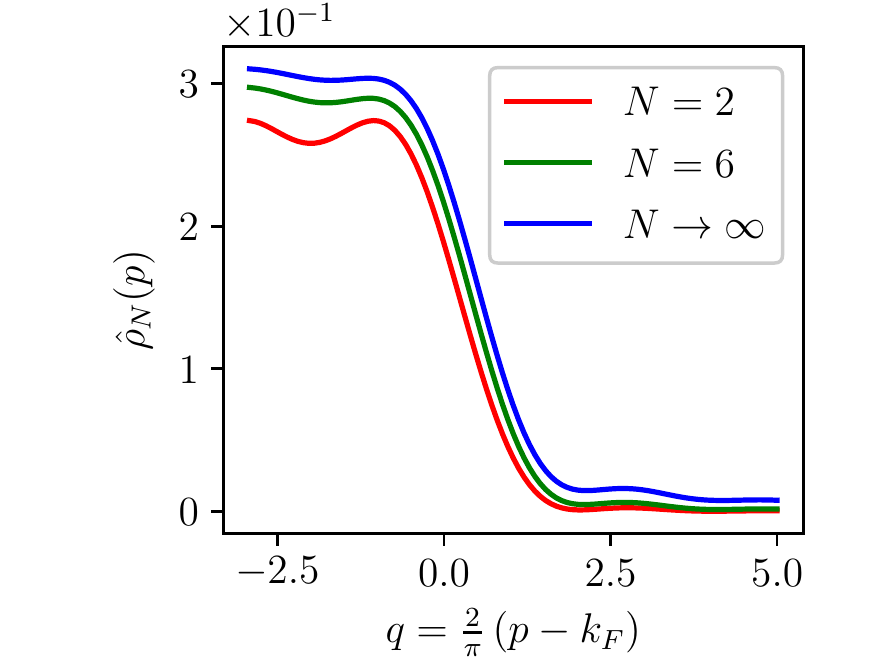}%
}\hfill
 \subfloat[Scaling function $\hat F_1(q)$ (\ref{rhop_2}).]{%
   \includegraphics[width=0.45\textwidth]{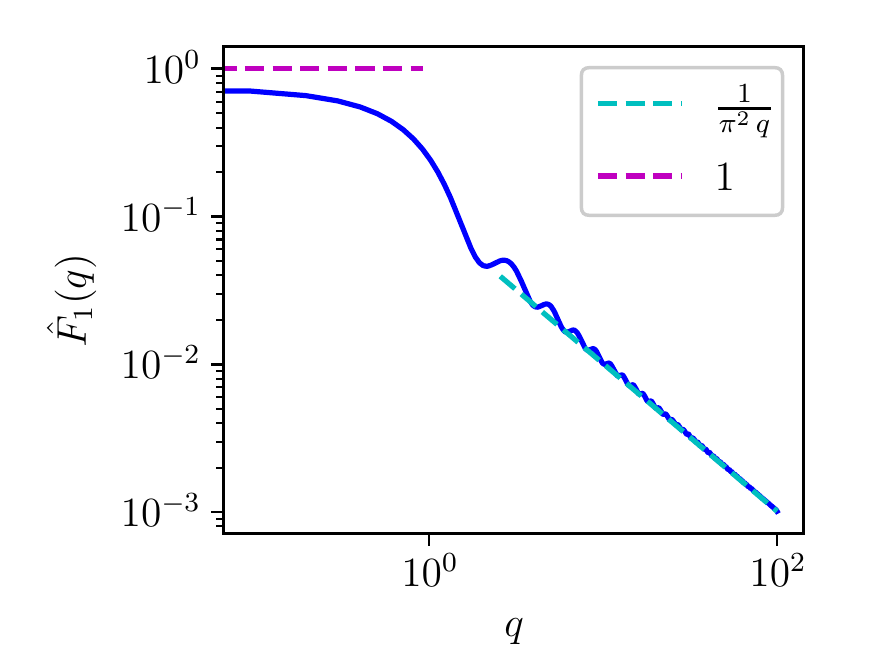}%
 }\hfill
 \caption{\textbf{a)} Exact and large $N$ description of the momentum density $\hat \rho_N(p)$ at the momentum edge $p=k_F+\frac{\pi}{2}\,q$ with $q=O(1)$. As $N$ increases, the momentum density approaches the scaling function $\hat F_1(q)$ given in (\ref{rhop_2}) (blue line). \textbf{b)} Log-log plot of the scaling function $\hat F_1(q)$ along with its asymptotic tails obtained in (\ref{eq:tailrhop}).}
 \label{fig:edgena}
\end{figure}

\begin{itemize}
\item[$\bullet$] (1) For $-k_F<p<k_F$: in this regime, the leading term of the density $\hat \rho_N(p)$ is easily obtained by integrating the Wigner function, as given in Eq. (\ref{rhop_1}) in the introduction. In this regime, at leading order for large $N$, the density is thus uniform, $\hat \rho_N(p) \simeq 1/\pi$. From the exact expression for $\hat \rho_N(p)$ in Eq. (\ref{density_p}), it is however possible to go beyond the leading order and obtain the first terms in the $1/N$ expansion, which show an intriguing dependence on the parity of $N$. Skipping some details, one obtains
\bea \label{rho_1overN}
\hat \rho_N(p) \simeq \frac{1}{\pi} - \frac{4}{\pi^3\, N} + 
\begin{cases}
&\dfrac{4}{\pi^3 N^2} \sin^2(p) + O(1/N^3) \quad, \quad {\rm if}\; N \; {\rm is \; even} \;, \\
& \\
&\dfrac{4}{\pi^3 N^2} \cos^2(p) + O(1/N^3) \quad, \quad {\rm if}\; N \; {\rm is \; odd}\;. \\
\end{cases}
\eea

\item[$\bullet$] (2) For $p$ close to $k_F$, with $p-k_F = O(1)$: in this regime, we start from the exact formula for the density in Eq. (\ref{density_p}) and set $p = (\pi/2)(N + q)$. We get
 \begin{align} \label{rho_21}
          \hat\rho_N\left(p = \frac{\pi}{2}(N + q)\right) =& \frac{1}{\pi^3}\sum_{k=1}^N\frac{4   k^2 \left((-1)^{k+1+N} \cos (\pi  q)+1\right)}{\left( k^2- (N+ q)^2\right)^2}\,,
    \end{align}
where we have used $\cos(N \pi + \pi q) = (-1)^N \cos{(\pi q)}$. In the limit of large $N$, the sum over $k$ in (\ref{rho_21}) is dominated by large $k$, with $k = O(N)$ and we thus
perform the change of variable $m= N-k$ in the sum and expand the summand to leading order for large $N$. This yields 
\bea \label{rho_22}
\hat\rho_N\left(p = \frac{\pi}{2}(N + q)\right) \simeq \frac{1}{\pi^3} \sum_{m=0}^{N-1} \frac{(-1)^{m+1}\cos(\pi q) + 1}{(m+q)^2}  \simeq \frac{1}{\pi^3} \sum_{m=0}^{\infty} \frac{(-1)^{m+1}\cos(\pi q) + 1}{(m+q)^2} \quad, \quad {\rm as} \quad N \to \infty \;.
\eea
Note that, using the identity $(-1)^{m+1} \cos{(\pi q)} + 1 = - \cos{(\pi q + \pi m)} + 1=2 \sin^2(\pi/2(m+q))$ for integer $m$, the last sum in (\ref{rho_22}) can also be written as
\bea \label{rho_23}
\hat\rho_N\left(p = \frac{\pi}{2}(N + q)\right) \simeq \frac{2}{\pi^3} \sum_{m=0}^\infty \frac{\sin^2{\left( \frac{\pi}{2}(m+q)\right)}}{(m+q)^2} \;,
\eea
whose structure is rather familiar in the theory of determinantal point processes (see also below). The last sum can eventually be expressed in terms of the tri-gamma function yielding the result given in Eq. (\ref{rhop_2}). The asymptotic behaviors of the scaling function $\hat F_1(q)$ for $q \to \pm \infty$ can be obtained from the ones for the tri-gamma function
\bea \label{asympt_psi1}
\psi^{(1)}(z) =
\begin{cases}
&\pi^2(1 + \cot^2(z)) + \dfrac{1}{z} + \dfrac{1}{2z^2} + O(1/z^3) \;, \; z \to - \infty \\
& \\
& \dfrac{1}{z} + \dfrac{1}{2z^2} + O(1/z^3)  \;, \quad\quad\quad\quad\quad\quad\quad\quad\; z \to + \infty  \;.
\end{cases}
\eea  
This yields, by injecting these asymptotic behaviors (\ref{asympt_psi1}) in (\ref{rhop_2}),
\bea
\hat F_1(q) = 
\begin{cases}
&1 + \dfrac{1}{\pi^2 q} + \dfrac{1}{\pi^2 q^2} \sin^2\left( \dfrac{\pi q}{2}\right) + O(1/q^3) \;, \; q \to - \infty \label{eq:tailrhop2} \\
& \\
& \dfrac{1}{\pi^2 q} + \dfrac{1}{\pi^2 q^2} \sin^2\left( \dfrac{\pi q}{2}\right) + O(1/q^3) \;, \; \quad\quad q \to + \infty \;.\label{eq:tailrhop}
\end{cases}
\eea 
In the limit $q \to -\infty$, the behavior in the first line in Eq. (\ref{eq:tailrhop2}) indicates that $\hat F_1(q)$ smoothly matches with the uniform density profile in the bulk, i.e. with the first term in Eq. (\ref{rho_1overN}), albeit with a slow
algebraic decaying correction. A similar slow algebraic decay is observed in the limit $q \to \infty$ [see the second line in Eq. (\ref{eq:tailrhop})]. On both sides, i.e. for $q \to \pm \infty$, the oscillations are only visible in the next-to-leading corrections, namely of order $O(1/q^2)$. Finally a plot of this function $\hat F_1(q)$ given in (\ref{rhop_2}) is shown in Fig.~\ref{fig:edgena}.
 

 \begin{figure}[htb]
\subfloat[Momentum density $\hat \rho_N(p)$ (\ref{density_p}).]{%
 \includegraphics[width=0.45\textwidth]{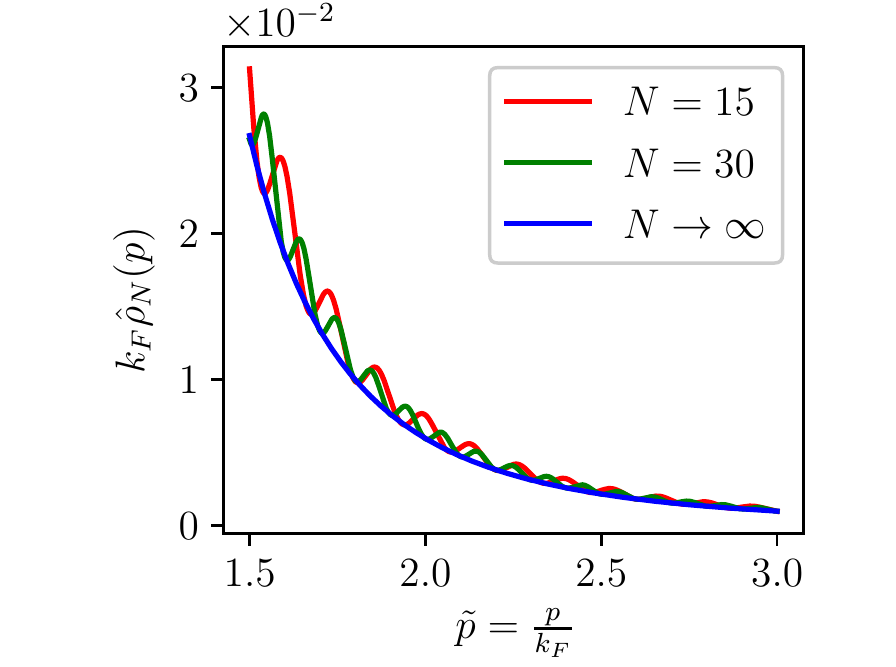}%
}\hfill
 \subfloat[Scaling function $\hat{\sf F}_1(\tilde p )$ (\ref{rhop_3}).]{%
   \includegraphics[width=0.45\textwidth]{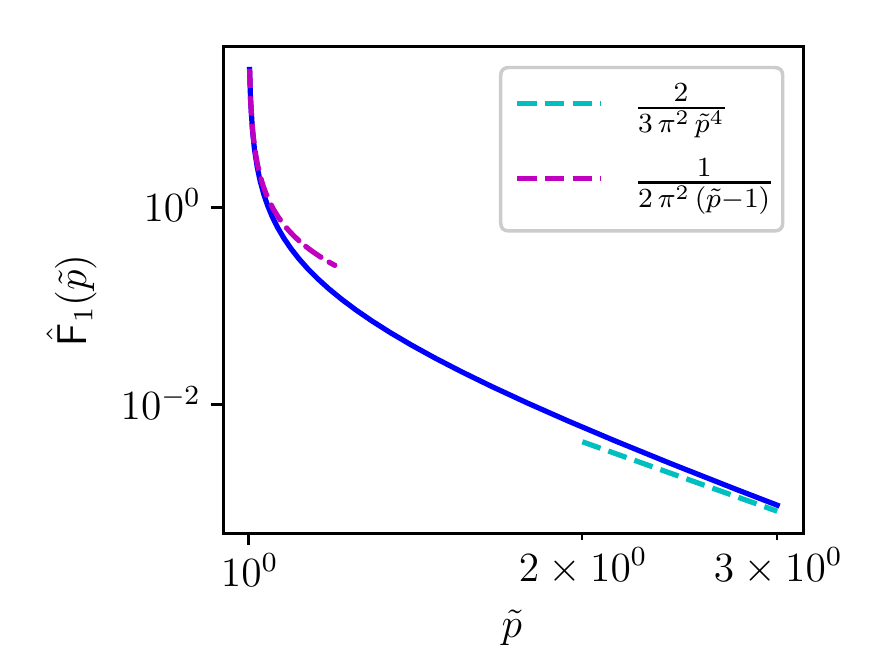}%
 }\hfill
 \caption{\textbf{Left panel:} Exact and large $N$ description of the momentum density $\hat \rho_N(p)$ for large momentum $p=\tilde p\,k_F$ with $\tilde p>1$. As $N$ increases, the momentum density approaches the scaling function $\frac{1}{k_F}\hat{\sf F}_1(\tilde p )$ given in (\ref{rhop_3}) (blue line). \textbf{Right panel:} Logarithmic plot of the scaling function $\hat{\sf F}_1(\tilde p )$ along with its asymptotic tails obtained in (\ref{asympt_rhop_3}).}
\label{fig:edgena2}
\end{figure}

\item[$\bullet$] (3) For $p= \tilde p \, k_F$, with $\tilde p > 1$: in this case, we start from the exact formula given in Eq. (\ref{density_p}) with $p= \tilde p \, k_F = \tilde p N \pi/2$. Since $\tilde p>1$, the denominator of the summand, i.e. $(\pi^2 k^2 - 4p^2)^2 = (\pi^2(k^2-\tilde p^2 N^2))^2$ does not vanish, since $k\leq N$ and therefore the sum over $k$ can be safely split into two terms
\bea \label{split}
\hat \rho_N\left(p=\tilde p \,\frac{N \pi}{2}\right) &=& \sum_{k=1}^N \frac{4 \pi  k^2 \left((-1)^{k+1} \cos (\tilde p N \pi)+1\right)}{\left(\pi ^2 k^2- N^2 \pi^2 \tilde p^2\right)^2} \nn \\ 
&=& \sum_{k=1}^N \frac{4 \pi  k^2}{\left(\pi ^2 k^2- N^2 \pi^2 \tilde p^2\right)^2} + \cos{(\tilde p N \pi)} \sum_{k=1}^N \frac{4 \pi k^2 \, (-1)^{k+1}}{\left(\pi ^2 k^2- N^2 \pi^2 \tilde p^2\right)^2}\,.
\eea
In the limit $N \to \infty$, keeping $\tilde p > 1$ fixed, one can show that the second term in (\ref{split}) is subleading compared to the first one, because of the alternating factor $(-1)^{k+1}$, and so
\bea \label{split2}
\hat \rho_N\left(p=\tilde p \,\frac{N \pi}{2}\right) \simeq \sum_{k=1}^N \frac{4 \pi  k^2}{\left(\pi ^2 k^2- N^2 \pi^2 \tilde p^2\right)^2} \;.
\eea
In the large $N$ limit, the discrete sum over $k$ can be replaced by an integral, which can be performed explicitly yielding the result given in Eq. (\ref{rhop_3}). Note that the associated scaling function ${\sf F}_1(\tilde p)$ does not exhibit any oscillatory behavior at all. Its asymptotic behaviors are straightforwardly obtained from the explicit expression in (\ref{rhop_3}) as
\bea  \label{asympt_rhop_3}
\hat{\sf F}_1(\tilde p) = 
\begin{cases}
&\dfrac{1}{2\pi^2(\tilde p-1)} + \dfrac{1}{\pi^2} \log(\tilde p -1) + O(1) \quad, \quad \tilde p \to 1 \;, \\
& \\
& \dfrac{2}{3 \pi^2 \tilde p^4} + \dfrac{4}{5\pi^2 \tilde p^6} + O(1/\tilde p^8) \quad, \quad \hspace*{1.2cm} \tilde p \to \infty \;.  
\end{cases}
\eea
Note that the leading term in the first line in Eq.~(\ref{asympt_rhop_3}), i.e. $\hat \rho_N(p) \simeq k_F^{-1}\hat{\sf F}_1(\tilde p) \simeq 1/(2 \pi^2(\tilde p-1)) = 1/(2\pi^2(p-k_F))$ as $\tilde p \to 1$, i.e. $p \to k_F$, matches with the large $q$ asymptotic behaviour in the second line of (\ref{eq:tailrhop}), i.e. $\hat \rho_N(p) \simeq \hat F_1(q = (2/\pi)(p - k_F)) \simeq 1/(\pi^3 q) = 1/(2 \pi^2(p-k_F))$. Note also the interesting logarithmic subleading correction in the first line in (\ref{asympt_rhop_3}). Finally, one can check that the large $\tilde p$ asymptotic behaviour in the second line in Eq. (\ref{asympt_rhop_3}) matches with the large $p$ asymptotic behavior of the exact finite $N$ expression of $\hat \rho_N(p)$ in Eq. (\ref{tail_rhoN}). In Fig.~\ref{fig:edgena2} we show a plot of this scaling function ${\sf F}_1(\tilde p)$.  
\end{itemize}

Finally, note that the scaling function 
$\hat {\sf F}_1(\tilde p)$ can also be obtained by integrating the Wigner function ${\cal W}_{\rm III}(\tilde s, \tilde p)$ given in Eq. (\ref{region_III}), i.e.,
\bea \label{FWIII}
\hat {\sf F}_1(\tilde p) = \frac{1}{\pi} \int_0^{+\infty} d\tilde s \, {\cal W}_{\rm III}(\tilde s, \tilde p) \;.
\eea
The factor $1/\pi = 2\times 1/(2 \pi)$ comes from the fact that one needs to integrate the Wigner function close to $x=-1$ {\it and} $x=+1$ (both yielding the same contribution) to obtain the full momentum density for $p = O(k_F)$. There exists a similar ``sum rule'' that relates $\hat F_1(q)$ to ${\cal W}_{\rm II}(x,q)$, i.e. 
\bea \label{FWII}
\hat F_1(q) = \int_{-1}^{1} dx \, {\cal W}_{\rm II}(x,q)\;,
\eea
which can easily be checked by comparing the formulae (\ref{region_II}) -- integrated over $x \in (-1,1)$ -- and (\ref{rho_23}).

      \subsection{Kernel in momentum space in $d=1$}
      \label{sec:ker1d}

As mentioned above, the momenta $p_i$'s, with $i=1,2, \cdots, N$ of the $N$ fermions in the ground-state of the hard-box potential
form a determinantal point process which is fully characterized by the kernel which reads \cite{DMS2018}
 \begin{align}
    \hat K_N(p,p') =& \sum_{k=1}^{N} \hat\phi^*_k(p)\hat\phi_k(p')\, ,
  \end{align}
 where $\hat \phi_k(p)$'s are the eigenfunctions in momentum space given in (\ref{eq:wfmoma}). For fixed $N$, it evaluates to
  \begin{align}
     \hat K_N(p,p')=& \frac{8}{\pi}\sum_{k=1}^{N} \frac{(k\pi)^2}{((k\pi)^2-4p^2)((k\pi)^2-4p'^2)}\sin\left(\frac{k\pi}{2}-p\right)\sin\left(\frac{k\pi}{2}-p'\right)\, .
     \label{eq:KNp}
  \end{align}
In particular, the density in momentum space is given by $\hat \rho_N(p) = \hat K_N(p,p)$. Indeed, one can easily check that evaluating Eq. (\ref{eq:KNp}) at coinciding points
$p=p'$ yields back the expression for the density in Eq. (\ref{density_p}). In the following, we compute the large $N$ limiting form of the kernel in the three different regions $(1)$, $(2)$
and $(3)$ that we have identified in the density.       
      
\begin{itemize}
\item[$\bullet$] (1) For $-k_F<p, p'<k_F$: in this case, one can show that the limiting kernel is given by the expression given in Eq.~(\ref{eq:KNp}) setting $N \to \infty$. Using 
the trigonometric identity $2\sin(a)\sin(b)=\cos(a-b)-\cos(a+b)$, we rewrite $\hat K_N(p,p')$ for $-k_F<p, p'<k_F$ as
\bea \label{kernel_bulk1}
 \hat K_N(p,p') \simeq \frac{4}{\pi}\sum_{k=1}^{\infty} \frac{(k\pi)^2}{((k\pi)^2-4p^2)((k\pi)^2-4p'^2)}\left(\cos\left(p'-p\right) + (-1)^{k+1}\cos(p'+p)\right) \;.
\eea
This sum over $k$ can be evaluated explicitly using the identities 
    \begin{eqnarray}
      \frac{1}{\pi}\sum_{k=1}^\infty \frac{k^2}{(k^2-z^2)(k^2-z'^2)} &=&  \frac{z'\cot(\pi z')-z\cot(\pi z)}{2(z^2-z'^2)}\, ,  \label{id1} \\
      \frac{1}{\pi}\sum_{k=1}^\infty \frac{(-1)^{k+1}\, k^2}{(k^2-z^2)(k^2-z'^2)} &=& \frac{z {\rm cosec}{(\pi z)} - z' {\rm cosec}(\pi z')}{2(z^2 - z'^2)}
    \end{eqnarray}
to get    
  \begin{equation} \label{sine_kernel}
 { \hat K_N\left(p,p'\right) \sim \frac{ \sin\left(p-p'\right)}{\pi(p-p')}} \;,
\end{equation}
which is the celebrated sine-kernel, well known in random matrix theory. Note that the typical scale momentum scale in this regime is $p = O(1/R)$ (with $R$ set to $R=1$ here), while the usual sine-kernel in position space occurs on microscopic scales of order $O(1/k_F)$. 

\item[$\bullet$] (2) For $p$ and $p'$close to $k_F$, with $p-k_F = O(1)$ and $p'-k_F = O(1)$,  setting $p= k_F+\frac{\pi}{2} q=\frac{\pi}{2}\left( N+ q\right)$ and $p'= k_F+\frac{\pi}{2} q'=\frac{\pi}{2}\left( N+ q'\right)$ in  (\ref{eq:KNp}) we get
   \begin{equation} \label{Kp2}
    \hat K_N(k_F+\frac{\pi}{2} q, k_F+\frac{\pi}{2} q') =\frac{8}{\pi^3}\sum_{k=1}^{N} \frac{k^2}{(k^2-(N+ q)^2)(k^2-(N+ q')^2)}\sin\left(\frac{\pi}{2}(k- q-N)\right)\sin\left(\frac{\pi}{2}(k- q'-N)\right) \,.
  \end{equation} 
By performing the change of variable $m = N-k$ in the sum we obtain
\bea \label{Kp3}
  \hat K_N(k_F+\frac{\pi}{2} q, k_F+\frac{\pi}{2} q') =\frac{8}{\pi^3}\sum_{m=0}^{N-1} \frac{(N-m)^2\, \sin\left(\frac{\pi}{2}(m+q)\right)\sin\left(\frac{\pi}{2}(m+q')\right)}{(m+q)(m+q')(2N-m+q)(2N-m+q')}\;.
\eea 
Finally, taking the large $N$ limit of the summand and sending the upper limit of the sum $N \to \infty$ yields the large $N$ limit of the kernel in this regime
\bea \label{Kp4}
\lim_{N \to \infty}  \hat K_N(k_F+\frac{\pi}{2} q, k_F+\frac{\pi}{2} q') = \frac{2}{\pi^3} \sum_{m=0}^\infty \frac{\sin\left(\frac{\pi}{2}(m+q)\right)}{(m+q)} \frac{\sin\left(\frac{\pi}{2}(m+q')\right)}{(m+q')} \;.
\eea
Note that this form (\ref{Kp4}) is reminiscent of the form of the kernels found for multi-critical fermions in a potential $V(x) \sim x^{2n}$ in the limit $n \to \infty$ with continuum integrals replaced by discrete sums \cite{DMS2018}. This sum over $m$ can be evaluated explicitly, leading to 
 \begin{equation} \label{kernel_kf1}
     {
      \begin{split}
      \lim_{N \to \infty}  \hat K_N(k_F+\frac{\pi}{2} q, k_F+\frac{\pi}{2} q')  =\frac{1}{\pi^3}\bigg[&\cos\left(\frac{\pi}{2}( q- q')\right)\zeta( q, q')\\
       +& \cos\left(\frac{\pi}{2}( q+ q')\right)\frac{1}{2}\bigg(\zeta\left(\frac{ q+1}{2},\frac{ q'+1}{2}\right)-\zeta\left(\frac{ q}{2},\frac{ q'}{2}\right) \bigg) \bigg]\,.
       \end{split}}
  \end{equation}
  where $\zeta(x,x')=(\psi ^{(0)}(x)-\psi ^{(0)}\left(x'\right))/(x-x')$. In particular, one has $\lim_{x'\rightarrow x}\zeta(x,x') = \psi^{(1)}(x)$. 
  Note that if one sets $q=q'$ in this expression (\ref{Kp4}), we recover the expression of the scaling function for the density in this regime (2) given in Eq. (\ref{rho_23}), as we should.

\item[$\bullet$] (3) For $p= \tilde p \, k_F$, and $p' = \tilde p'\, k_F$ with $\tilde p > 1$:  Setting $p=\tilde p k_F=\tilde p \left(\frac{N\pi}{2}\right)$ and $p'= \tilde p' k_F=\tilde p' \left(\frac{N\pi}{2}\right)$ in  (\ref{eq:KNp}) gives
\bea \label{Kp3_1}
&&\hat K_N\left(p=\frac{N\pi}{2}\tilde p, p'= \frac{N\pi}{2} \tilde p'\right) \nonumber \\
&&= \frac{4}{N^4 \pi^3} \sum_{k=1}^N \frac{k^2}{\left(\left(\frac{k}{N}\right)^2 - \tilde p^2 \right)\left(\left(\frac{k}{N}\right)^2 - \tilde p'^2 \right)}\left(\cos{\left[(\tilde p' - \tilde p)\frac{N \pi}{2}\right]} + (-1)^{k+1} \cos{\left[(\tilde p' + \tilde p)\frac{N \pi}{2}\right]}\right) \;.
\eea
In the limit of large $N$, one can show that the first term in (\ref{Kp3_1}), i.e. $\propto \cos{\left[(\tilde p' - \tilde p)\frac{N \pi}{2}\right]}$ dominates the second term $\propto  (-1)^{k+1} \cos{\left[(\tilde p' + \tilde p)\frac{N \pi}{2}\right]}$ because of the alternating sign of the latter. Hence as $N \to \infty$ it is natural to consider the scaling limit where 
\bea \label{scaling_}
(\tilde p' - \tilde p)\frac{N \pi}{2} = z 
\eea
is finite. Note that this corresponds to a limit where $\tilde p' - \tilde p = O(1/k_F) = O(1/N)$. Therefore, from Eq. (\ref{Kp3}) one gets in this scaling limit, keeping $z$ fixed (and at leading order for large $k_F$)
\bea \label{Kp3_2}
\hat K_N\left(p=\frac{N\pi}{2}\tilde p, p'= \frac{N\pi}{2} \tilde p'\right) \simeq \frac{1}{k_F} {\sf F_1}(\tilde p) \cos{z} \;,
\eea
where the function ${\sf F}_1(\tilde p)$ is given in Eq. (\ref{rhop_3}). Here also, if we set $p=p'$ in this expression (\ref{Kp3_2}), one recovers the expression for the density given in Eqs. (\ref{rhop_3}) and (\ref{split2}). 
\end{itemize}

    \section{Wigner function for $d>1$}\label{sec:wignerd}


\begin{figure}[t]
  \begin{center}
    \includegraphics[width=0.35\textwidth]{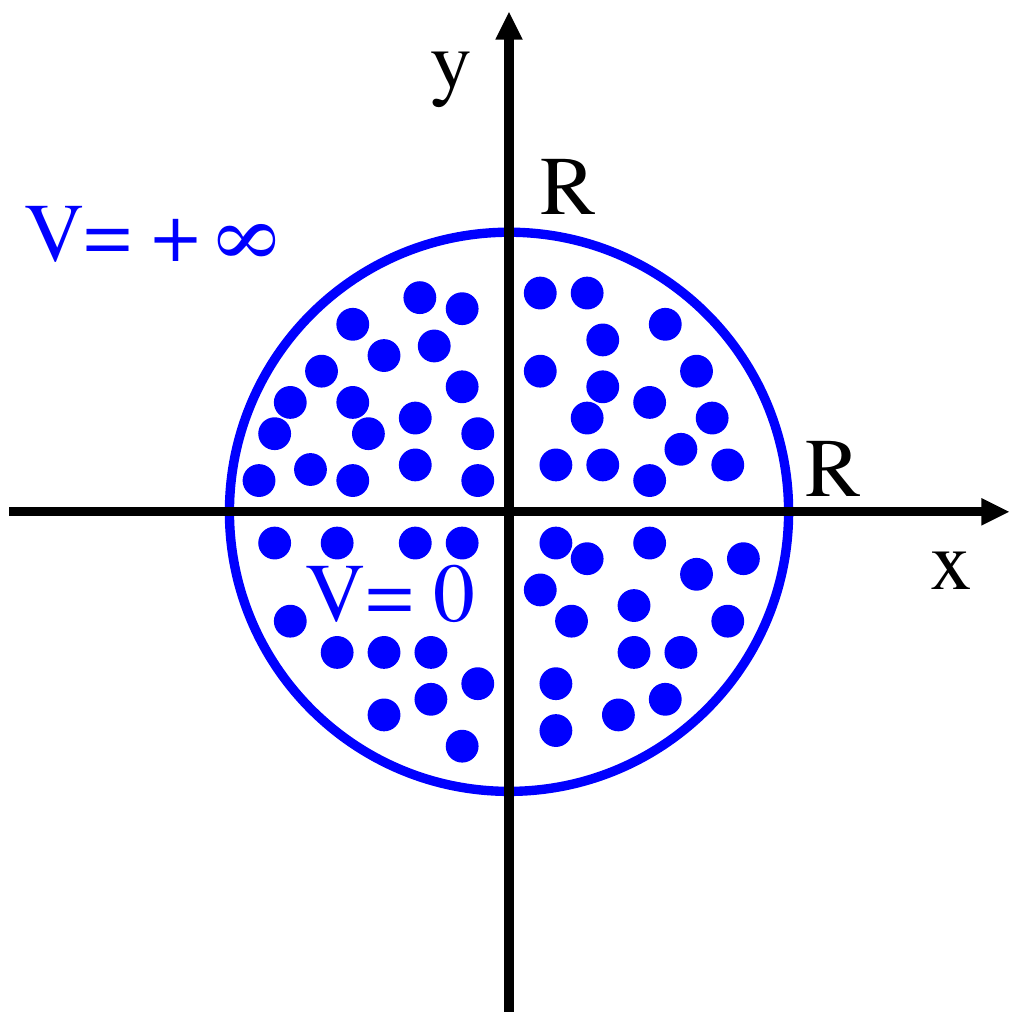}
    \caption{Schematic representation of a $d$-dimensional spherical box of radius $R$ in $d=2$ dimensions (\ref{def_V_Box}). The potential is zero inside the box, $V(x,y)=0$ for $x^2+y^2<R^2$, and infinite outside of it, $V(x,y)=+\infty$ for $x^2+y^2\geq R^2$.}
    \label{fig:2dpotential}
  \end{center}
\end{figure}

We now consider the case of $N$ fermions in a $d$-dimensional spherical hard box potential (\ref{def_V_Box}) -- see also Fig. \ref{fig:2dpotential} -- in their ground state. Here also we set the radius of the box to unity, i.e. $R=1$. In this case, as we did in the one-dimensional case, it is convenient to write the Wigner function in terms of the $d$-dimensional kernel $K_N(\bf x, \bf y)$ \cite{DDMS2018}, as in Eq. (\ref{W_vs_Kernel})
\bea \label{start_d_dim}
W_N({\bf x}, {\bf p}) = \frac{1}{(2 \pi)^d} \int d {\bf y} \, e^{i {\bf p\cdot  y}} K_N\left({{\bf x} - \frac{\bf y}{2}}, {{\bf x} + \frac{\bf y}{2}} \right) \;.
\eea
Since the eigenfunctions vanish outside the box, and so does the kernel,  the domain of integration over ${\bf y}$ in (\ref{start_d_dim}) is
\bea  \label{domain}
\Big |  {\bf x} - \frac{\bf y}{2} \Big | \leq 1 \;\; \& \;\;   \Big | {\bf x} + \frac{\bf y}{2}  \Big | \leq 1 \;.
\eea
In this case, the kernel $K_N(\bf x, \bf y)$ can be explicitly computed -- see Eq. (95) of \cite{lacroix2018non} -- but the resulting expression is rather complicated
and this would lead, once inserted in Eq. (\ref{start_d_dim}), to a quite cumbersome expression of the Wigner function, whose full asymptotic analysis for large $N$ goes beyond the scope of the present paper. 

To study the large $N$ limit, let us instead start with the LDA prediction in Eq. (\ref{LDA}). This formula immediately tells us that for the $d$-dimensional spherical hard box potential (\ref{def_V_Box}) the Fermi surf is the product of two $d$-dimensional spheres defined by $|{\bf x}| = 1$ and $|{\bf p}| = k_F$ in position. Inside the Fermi surf, the Wigner function is constant $W_N({\bf x}, {\bf p}) \approx \frac{1}{(2 \pi)^d}$ [see Eq. (\ref{LDA_ddim})] while $W_N({\bf x}, {\bf p})$ vanishes outside the Fermi surf. Note that this prediction from the LDA can be obtained in more controlled way by starting from the exact expression for the Wigner function in (\ref{start_d_dim}) and using the large $N$ limiting form of the kernel $K_N({\bf x}, {\bf y})$ in the bulk, i.e., far from the wall. We refer the reader to Ref. \cite{DDMS2018} for more details on this computation of the Wigner function far from the Fermi surf.

\begin{figure}
  \begin{center}
    \includegraphics[width=0.30\textwidth]{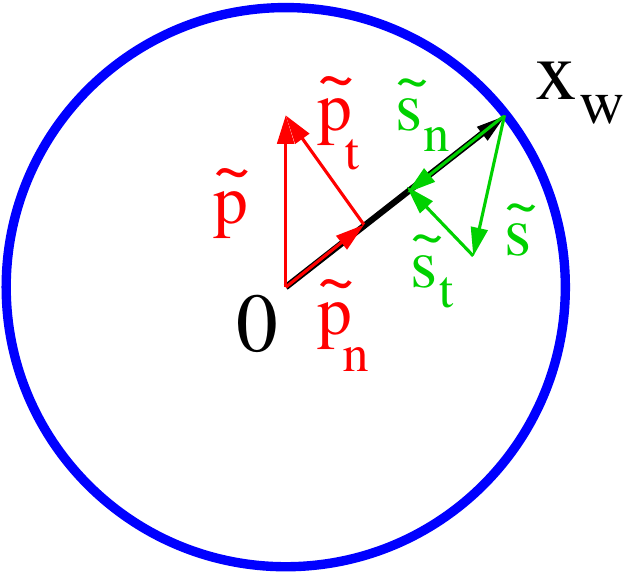}
    \caption{Schematic representation of the rescaled position vector $\mathbf{x}={\bf x}_w + k_F^{-1} {\bf \tilde  s}$ and momentum vector $\mathbf{p}=k_F {\bf \tilde p}$ in a $d$-dimensional box (\ref{radial4}). The blue circle represents the edge of the hard box (as in  Fig. \ref{fig:2dpotential}). The subscript $n$ refers to the component of the vector that is parallel to $\mathbf{x}_w$ and the subscript $t$ refers to the component that is perpendicular to $\mathbf{x}_w$. }
    \label{fig:2dnt}
  \end{center}
\end{figure}

Instead, we restrict our study of the Wigner function $W_N(\bf x, \bf p)$ to near the wall, i.e. the analogue of the regime III in the one-dimensional case (see Fig. \ref{fig:fermisurf}).   
We thus set, adopting the notations of Ref. \cite{lacroix2018non} (see Fig. \ref{fig:2dnt})
\bea \label{def_var}
{\bf x} = {\bf x}_w + k_F^{-1} {\bf \tilde  s} \quad, \quad {\bf p} = k_F \, {\bf \tilde p}
\eea
where ${\bf x}_w$ labels a point exactly at the wall, hence such that $| {\bf x}_w| = 1$. For large $N$, and for $|{\bf p}| = O(k_F)$ the integral over ${\bf y}$ in Eq. (\ref{start_d_dim}) is dominated by $| {\bf y} | = O(k_F^{-1})$. Therefore, we perform the change of variable ${\bf \tilde y} = k_F {\bf y}$, leading to
\bea  \label{W_d_dim_1}
W_N({\bf x}_w + k_F^{-1} {\bf \tilde  s},  k_F {\bf \tilde p}) = \frac{1}{(2 \pi)^d} \int d {\bf \tilde y}  \, e^{i {\bf \tilde p\cdot  \tilde y}} \frac{1}{k_F^d}K_N\left({\bf x}_w + \frac{1}{k_F}({\bf \tilde  s}- \frac{\bf \tilde y}{2}), {\bf x}_w + \frac{1}{k_F}({\bf \tilde  s} + \frac{\bf \tilde y}{2})\right) \;.
\eea
Following Ref. \cite{lacroix2018non}, we denote by ${\bf u}_t$ and ${u_n}$, respectively the transverse and the normal component of an arbitrary vector ${\bf u}$ (see Fig. \ref{fig:2dnt} where ${\bf u}$ can represent either $\tilde{\bf s}$ or $\tilde{\bf p}$). In the large $N$ limit, we can then use the limiting form of the kernel near the wall, i.e., {\it near the edge} of the Fermi gas 
\bea \label{kernel_d_largeN}
\frac{1}{k_F^d} K_N\left({\bf x}_w + \frac{1}{k_F}({\bf \tilde  s}- \frac{\bf \tilde y}{2}), {\bf x}_w + \frac{1}{k_F}({\bf \tilde  s} + \frac{\bf \tilde y}{2})\right) \underset{N \to \infty}{\longrightarrow} K_d^{e} \left( {\bf \tilde  s}- \frac{\bf \tilde y}{2}, {\bf \tilde  s}+ \frac{\bf \tilde y}{2}\right)
\eea
where the edge kernel $K_d^e$ was computed in \cite{lacroix2018non}. Note also that the domain of integration for ${\bf y}$ in (\ref{domain}) translates into the following domain for ${\bf \tilde y}$ (in the limit $N \to \infty$, or equivalently $k_F \to \infty$)
\bea \label{domain_tilde}
-2 \tilde  s_n \leq \tilde y_n \leq 2 \tilde  s_n \;, \; {\bf \tilde y}_t \in {\mathbb R}^{d-1} \;.
\eea
In Ref. \cite{lacroix2018non} different representations of the edge kernel were obtained. Here we present a computation of the Wigner function at the edge, 
obtained by substituting the scaling form (\ref{kernel_d_largeN}) in (\ref{W_d_dim_1}) and then using a ``radial'' representation of $K_d^e$. In the Appendix \ref{app:Bessel}, we provide an alternative derivation using a representation of this kernel in terms of Bessel functions, yielding of course to the same result. 

A useful representation of the hard wall edge kernel is (see Eq. (128) in \cite{lacroix2018non})
\bea\label{radial}
&&K_d^{e}({\bf u}, {\bf   v}) = \int_{{|{\bf l}| < 1}} \frac{d^{d-1} {\bf l}}{(2 \pi)^{d-1}} e^{i {\bf l \cdot (u_t -   v_t)}} \sqrt{1 - {\bf l}^2} K_1^e(u_n \sqrt{1 - {\bf l}^2},  v_n \sqrt{1-{\bf l}^2}) \\
&& K_1^e(x,y) = \frac{\sin(x-y)}{\pi(x-y)} - \frac{\sin(x+y)}{\pi(x+y)} \;. \label{1d} 
\eea
Inserting (\ref{radial}) and (\ref{1d}) in Eq. (\ref{W_d_dim_1}), using (\ref{kernel_d_largeN}), one finds
\bea
&&W_N({\bf x}_w + k_F^{-1} {\bf \tilde  s},  k_F {\bf \tilde p}) \nonumber \\
&&\underset{N \to \infty}{\longrightarrow}\frac{1}{(2 \pi)^d} \int d^{d-1} {\bf \tilde y}_t \int_{-2 \tilde  s_n}^{2 \tilde  s_n}  d\tilde y_n \, e^{i {\bf \tilde p_t \cdot \tilde y_t} + i \tilde p_n \tilde y_n} \int_{{|{\bf l}| < 1}} \frac{d^{d-1} {\bf l}}{(2 \pi)^{d-1}} \, e^{-i {\bf l \cdot \tilde y_t}} \sqrt{1 - {\bf l}^2} \nonumber \\
&& \times \, K_1^e\left(\left(\tilde  s_n-\frac{\tilde y_n}{2}\right) \sqrt{1 - {\bf l}^2},\left(\tilde  s_n+\frac{\tilde y_n}{2}\right) \sqrt{1 - {\bf l}^2}\right) \label{radial2} \;.
\eea
Under this form (\ref{radial2}), we see that the integral over ${\bf \tilde y}_t$ can be performed straightforwardly, yielding simply $(2 \pi)^{d-1}~\delta({\bf \tilde p}_t ~- ~{\bf l})$. Thus one obtains
\bea \label{radial3}
&&W_N({\bf x}_w + k_F^{-1} {\bf \tilde  s},  k_F {\bf \tilde p}) \underset{N \to \infty}{\longrightarrow}\frac{\Theta(1-{\bf \tilde p_t}^2)}{(2 \pi)^d} \sqrt{1-{\bf \tilde p_t}^2}\label{eq:dintdelta}\\
&&\times\int_{-2 \tilde  s_n}^{2 \tilde  s_n}  d\tilde y_n \, e^{ i \tilde p_n \tilde y_n}\, K_1^e\left(\left(\tilde  s_n-\frac{\tilde y_n}{2}\right) \sqrt{1 - {\bf \tilde p}_t^2},\left(\tilde  s_n+\frac{\tilde y_n}{2}\right) \sqrt{1 - {\bf \tilde p}_t^2}\right) \;. \nonumber
\eea
Finally, performing the integral over $\tilde y_n$, we find that the Wigner function $W_N({\bf x},{\bf p})$ for the spherical hard box takes at large $N$ the following scaling form, 
which is our main result in dimension $d$
\bea \label{radial4}
&&W_N({\bf x}_w + k_F^{-1} {\bf \tilde  s},  k_F {\bf \tilde p}) \approx \frac{1}{(2 \pi)^d} {\sf W}_{\rm III}(\tilde {\bf s}, \tilde {\bf p} ) \\
&&{\sf W}_{\rm III}(\tilde {\bf s}, \tilde {\bf p} ) =\frac{\Theta(1-{\bf \tilde p_t}^2)}{\pi}\Bigg[ {\rm Si}\left(2 \tilde  s_n(\sqrt{1-{\bf \tilde p}_t^2}+\tilde p_n)\right) + {\rm Si}\left(2 \tilde  s_n(\sqrt{1-{\bf \tilde p}_t^2} - \tilde p_n)\right)- \frac{\sin{(2 \tilde  s_n \sqrt{1 - {\bf \tilde p}_t^2})} \sin{(2 \tilde  s_n \tilde p_n)}}{{\tilde p}_n {\tilde  s}_n}\Bigg] \;, \nonumber
\eea
where we recall that ${\rm Si}(x)=\int_0^x \sin(t)/t\, dt$ and the notations ${\bf \tilde p_t}$, ${\bf \tilde s_t}$, $\tilde p_n$ and $\tilde s_n$ 
are defined in the Fig. \ref{fig:2dnt}.
Remarkably, the form of the scaling function ${\sf W}_{\rm III}(\tilde {\bf s}, \tilde {\bf p} )$ is independent of the dimension $d$. In particular, in the case $d=1$, one has ${\bf \tilde p}_t = 0$ and one can check that ${\sf W}_{\rm III}(\tilde {\bf s}, \tilde {\bf p} ) = {\cal W}_{\rm III}(\tilde s_n, \tilde p_n)$ given in (\ref{region_III}), as it should. The generic structure of this result for the Wigner function in regime III has an interesting semi-classical interpretation that we discuss in the Appendix \ref{app:single}. 

As in the one-dimensional case, one can show (see Appendix \ref{app:single-hardwall_d}) that the limiting Wigner function in (\ref{radial4}) corresponds to the Wigner function for non-interacting fermions on a semi-infinite space $x_d > 0$ -- we recall that we use the notation ${\bf x} = (x_1, x_2, \cdots, x_d)$ -- in the presence of a $d$-dimensional hard wall potential of the form   
\bea \label{hw_ddim}
V(\bf x) = \begin{cases}
&+ \infty \;, \; x_d<0\;, \\
&0 \; , \; \hspace*{0.5cm} x_d > 0 \;. \\
\end{cases}
\eea

\section{Conclusions and perspectives}\label{sec:conclusion}

In this paper, we have studied the Wigner function $W_N({\bf x}, {\bf p})$ for $N$ noninteracting fermions in a $d$-dimensional spherical hard box of radius $R$ at temperature $T=0$, going far beyond the prediction of the LDA (\ref{LDA}). 
In particular, we have shown that, near the ``Fermi surf'' (see Fig. \ref{fig:fermisurf}), the Wigner function exhibits an edge behavior in the large $N$ limit which is quite different from the one found previously for smooth potentials \cite{balazs,Wiegman,DDMS2018}.  
For ${\bf x}$ close to the wall (regime III in Fig. \ref{fig:fermisurf}), we have computed explicitly the scaling function describing $W_N({\bf x}, {\bf p})$ and found, rather remarkably, that it is independent of the space dimension $d$. It is quite different from the scaling function (of the Airy type) which describes the Wigner function at the edge for a smooth potential.

Focusing on $d=1$, we were able to derive a more complete description of the Wigner function everywhere along the Fermi surf as explained in Fig. \ref{fig:fermisurf}. We have computed explicitly three nontrivial scaling functions along the Fermi surf. 
Finally, in $d=1$ we were also able to compute explicitly the momentum distribution $\hat \rho_N(p)$ of the fermions for all $p$ and $N$. This momentum distribution, for large $N$, exhibits a remarkable algebraic tail for $p \gg k_F$, i.e., $\hat \rho_N(p) \propto 1/p^4$. This is very different from the corresponding tail of the momentum distribution for fermions in a smooth potential, where it has typically a super-exponential tail \cite{DMS2018}.  However, this $1/p^4$ tail is also reminiscent of the similar tail found in interacting quantum systems with contact repulsion.

A natural question is what happens if the infinite wall is replaced by a continuous singular potential of the type $V(x) \propto 1/x^\gamma$ with $\gamma>0$? In Ref. \cite{lacroix2018non} it was shown that, for $1 \leq \gamma < 2$, the kernel near the singularity is identical to that of a hard wall at $x=0$. Hence we expect that, for $1 \leq \gamma < 2$,  
the Wigner function will also be described by the same scaling function ${\cal W}_{\rm III}(\tilde s, \tilde p)$ as the hard wall case discussed in this paper. The special case $\gamma=2$ is discussed in Appendix \ref{app:single-hardwall-inv} where the result is different from the hard wall case, as expected. In view of recent works on finite square well potential \cite{step}, it would also be interesting to study the Wigner function in this case.  

Finally, in higher dimension $d>1$, we have focused on the behavior of the Wigner function when the position ${\bf x}$ is close to the wall, while $|{\bf p}| = O(k_F)$. As in $d=1$, it would be interesting to investigate the behavior of the Wigner function close to the momentum edge $|{\bf p}|-k_F = O(1/R)$ and also the distribution of the momentum. Another question is what happens at finite temperature? The finite temperature Wigner function near the Fermi surf is straightforward to compute using the formula (84) in Ref. \cite{DDMS2018} which relates the finite temperature Wigner function to its zero temperature counterpart.


\begin{thebibliography}{100}
\bibitem{footnote1}
Note that the Dirichlet kernel (\ref{DN}) corresponds to the kernel for $(2N+1)$ fermions on the unit circle in their (non-degenerate) ground state.
\bibitem{footnote2}
This can be shown by substituting the expression for $D_N(z)$ in Eq. (\ref{DN}) in the integral on the left hand side of Eq. (\ref{eq:diricheletDelta}), followed by a change of variable $(N+1/2)x=y$ and taking the $N \to \infty$ limit. We also use the identity $\int_{0}^\infty \sin{y}/y \,dy= \pi/2$.
\end{thebibliography}

    \section*{Acknowledgments}
    This work was partially supported by the Luxembourg National Research Fund (FNR) (App. ID 14548297) and by ANR grant ANR-17-CE30- 0027-01 RaMaTraF. We thank C. Salomon for useful discussions.

\begin{appendix}

\section{Wigner function for a single particle in hard-wall potentials and semi-classical interpretation}\label{app:single}

In this section, we briefly recall the Wigner function and its semi-classical interpretation for a single-particle in hard-wall potentials (see e.g. \cite{CKM1991}). We start with a single particle on the infinite line described by a single 
plane wave, i.e., 
\bea \label{PW}
\varphi_{\rm PW}(x) =  \frac{1}{\sqrt{2 \pi}}e^{i k x} \quad, \quad x \in {\mathbb R} \;,
\eea
where the subscript `PW' refers to 'plane wave'. In this case the single-particle Wigner function defined in Eq. (\ref{def_W1}) is given by
\bea \label{W_PW}
W_{\rm PW}(x,p) = \frac{1}{2\pi}\delta(p - \hbar k) \;.
\eea 
In this simple case, interpreting the Wigner function as a quasi-distribution in the phase space $(x,p)$, the result (\ref{W_PW}) is what
one would expect from a classical analogy. Indeed the state described by $\varphi_{\rm PW}(x)$ in (\ref{PW}) has a well defined
momentum $\hbar k$ -- since this is an eigenstate of the momentum operator $\hat p$ with eigenvalue $p = \hbar k$ -- and therefore
the corresponding Wigner function is $W_{\rm PW}(x,p)  \propto \delta(p - \hbar k)$. 

Let us now consider a superposition of two counter-propagating plane waves
 \bea \label{2PW}
\varphi_{\rm 2PW}(x) =  \frac{1}{\sqrt{\pi}}\sin{kx}  = \frac{1}{i \sqrt{2}}\frac{1}{\sqrt{2 \pi}} e^{i k x} -  \frac{1}{i \sqrt{2}}\frac{1}{\sqrt{2 \pi}}e^{-ikx}\quad, \quad x \in {\mathbb R}  \;,
\eea
such that now the state is a linear combination (with equal amplitude) of two states with momentum $\pm \hbar k$. By substituting this expression (\ref{2PW})
in  Eq. (\ref{def_W1}), it is straightforward to evaluate the Wigner function which reads
\bea \label{W_2PW}
W_{\rm 2PW}(x,p) =  \frac{1}{2\pi} \left( \frac{1}{2} \delta(p-\hbar k) +   \frac{1}{2} \delta(p +\hbar k) - \cos(2 k x) \delta(p) \right) \;.
\eea
The two first delta functions $\delta(p-\hbar k)$ and $\delta(p+\hbar k)$ can simply be understood, from the classical analogy, from the interpretation of the wave function in (\ref{2PW}) 
mentioned above, being a simple extension of (\ref{W_PW}). However, the third term $\propto \delta(p)$ does not have a classical analogue and is the result of ``quantum interferences'' between the two plane waves. 

Let us now consider the case where the particle is constrained to stay on the semi-infinite line with $x \geq 0$ and in the presence of a hard wall at the origin
\bea \label{V_flat_single_hw_app}
V(x) = 
\begin{cases}
&+ \infty \;, \; x<0\;, \\
&0 \; , \; \hspace*{0.5cm} x > 0 \;. \\
\end{cases}
\eea
Let us now consider an eigenstate 
\bea \label{phi_W}
\varphi_{\rm HW}(x) = \sqrt{\frac{2}{\pi}} \Theta(x) \sin{(k\,x)} \;,\; k > 0 \;,
\eea
where the subscript `HW' refers to 'hard wall'. It is similar to the superposition of the two plane waves considered above (\ref{2PW}) but now the particle is constrained to stay on the half line $x>0$. The Wigner function reads in this case
\bea \label{W_HW}
W_{\rm HW}(x,p) = \frac{1}{\pi} \left( \frac{1}{2} f_{\rm D}(x,p - \hbar k) + \frac{1}{2} f_{\rm D}(x,p + \hbar k)  - \cos{2 k x} f_{\rm D}(x,p)\right) \quad, \quad f_{\rm D}(x,p) = \frac{1}{\pi} \, \frac{\sin{\frac{2x}{\hbar}p}}{p} \;.
\eea
By comparing this result for the Wigner function in the presence of the wall (\ref{W_HW}) with the one obtained without the wall in Eq. (\ref{W_2PW}) we see that they have exactly the same structure except that the Dirac delta function of $p$ in (\ref{W_2PW}) is ``broadened'' by the presence of the wall and is replaced by an $x$-dependent function $f_{\rm D}(x,p)$. In fact $f_{\rm D}(x,p) \to \delta(p)$ far from the wall, i.e., as $x \to \infty$. 

Finally, we note that a similar structure (\ref{W_HW}) also holds for the Wigner function corresponding to an eigenstate of a single particle in a hard box $x \in [-1,1]$ [see Eq. (\ref{eq:psipsia2})]. Indeed, the expression in Eq. (\ref{eq:psipsia2}) -- where we have set $\hbar  =1$ -- can be written as in Eq. (\ref{W_HW}), up to a global prefactor, with the substitution $k \to n \pi/2$ and $x \to 1-|x|$, which is actually the distance to the nearest hard wall.

\section{Asymptotic analysis of ${\cal W}_{\rm II}(x, q)$ for large $|q|$}\label{app:WII}

In this section, we provide some details about the asymptotic analysis of ${\cal W}_{\rm II}(x, q)$ for large $|q|$. Our starting point is the formula in Eq. (\ref{eq:EII}) which we write as
\bea \label{WII_app1}
{\cal W}_{\rm II}(x,q) = \frac{1}{2} -  {\cal I}(\pi(1-x),q) \quad {\rm where} \quad {\cal I}(z,q) = \int_{0}^{z}  D_{q-1}(u)\,du = \frac{1}{2\pi} \int_0^z \frac{\sin\left[(q-\frac{1}{2})u\right]}{\sin(u/2)}\, du\;.
\eea
To analyse the function ${\cal I}(z,q)$ for large $|q|$ and $z>0$, it is convenient to write 
\bea \label{decompo}
\frac{1}{\sin(u/2)} = \frac{2}{u} + g(u) \quad , \quad g(u) =  \frac{1}{\sin(u/2)} - \frac{2}{u} \;.
\eea
As we will see, the advantage of this decomposition (\ref{decompo}) is that $g(u)$ is a smooth function near $u = 0$. Inserting (\ref{decompo}) in the definition of ${\cal I}(z,q)$ in (\ref{WII_app1}) we get
\bea \label{WII_app2}
 {\cal I}(z,q) = \frac{1}{\pi} {\rm Si}\left((q-\frac{1}{2})z\right) + \frac{1}{2\pi} \int_0^z du \, \sin\left[(q-\frac{1}{2})u\right]\, g(u) \;,
\eea
where we recall that ${\rm Si}(x)$ is the Sine-integral function ${\rm Si}(x) = \int_0^x (\sin t) /t \, dt$. Using its asymptotic behavior given in (\ref{Si_asympt}) one finds
\bea \label{WII_app2a}
 \frac{1}{\pi} {\rm Si}\left((q-\frac{1}{2})z\right) = \frac{1}{2} {\rm sgn}(q) - \frac{1}{\pi} \frac{\cos\left[(q-\frac{1}{2})z\right]}{q\,z} + O(1/q^2) \;,
\eea
where we have used that $z>0$. To obtain the large $q$ behavior of the integral over $u$ in (\ref{WII_app2}) we perform an integration by parts [i.e., deriving $g(u)$ and integrating $\sin((q-1/2)u)$], one gets
\bea \label{WII_app3}
\frac{1}{2\pi} \int_0^z du \, \sin\left[(q-\frac{1}{2})u\right]\, g(u) = \frac{1}{2\pi}\left( - \frac{g(z)}{q-1/2} \cos\left[(q-\frac{1}{2})z\right]+ \frac{1}{q-1/2} \int_0^z du\, g'(u) \cos \left[(q-\frac{1}{2})u\right]\right) \;,
\eea
where we have used $g(0)=0$. Since $g'(u)$ is a perfectly regular function near $u=0$ one can again perform an integration by parts which shows that the remaining integral in (\ref{WII_app3}) is of order 
$O(1/q^2)$. Hence to leading order for large $|q|$, we get
\bea \label{WII_app3a}
\frac{1}{2\pi} \int_0^z du \, \sin\left[(q-\frac{1}{2})u\right]\, g(u)  = - \frac{g(z)}{2\pi\,q} \cos\left[(q-\frac{1}{2})z\right] + O(1/q^2) \;.
\eea
Finally, inserting the asymptotic behaviours (\ref{WII_app2a}) and (\ref{WII_app3a}) in Eq. (\ref{WII_app2}) we obtain
\bea \label{WII_app4}
 {\cal I}(z,q)  =  \frac{1}{2} {\rm sgn}(q) - \frac{ \cos\left[(q-\frac{1}{2})z\right]}{2\pi q} \left(\frac{2}{z} + g(z) \right) + O(1/q^2) =  \frac{1}{2} {\rm sgn}(q) - \frac{1}{2\pi q}\frac{ \cos\left[(q-\frac{1}{2})z\right]}{\sin(z/2)} + O(1/q^2) \;.
\eea
Finally, inserting this expansion (\ref{WII_app4}) with $z = \pi(1-|x|)$ in Eq. (\ref{WII_app1}) one obtains the asymptotic expansions given in Eq. (\ref{largek_II}) in the text.

\section{Wigner function for a single hard wall in $d=1$} \label{app:single-hardwall}

\subsection{The case of a flat potential}\label{app:single-hardwall-flat}

\begin{figure}[t]
  \begin{center}
    \includegraphics[width=0.4\textwidth]{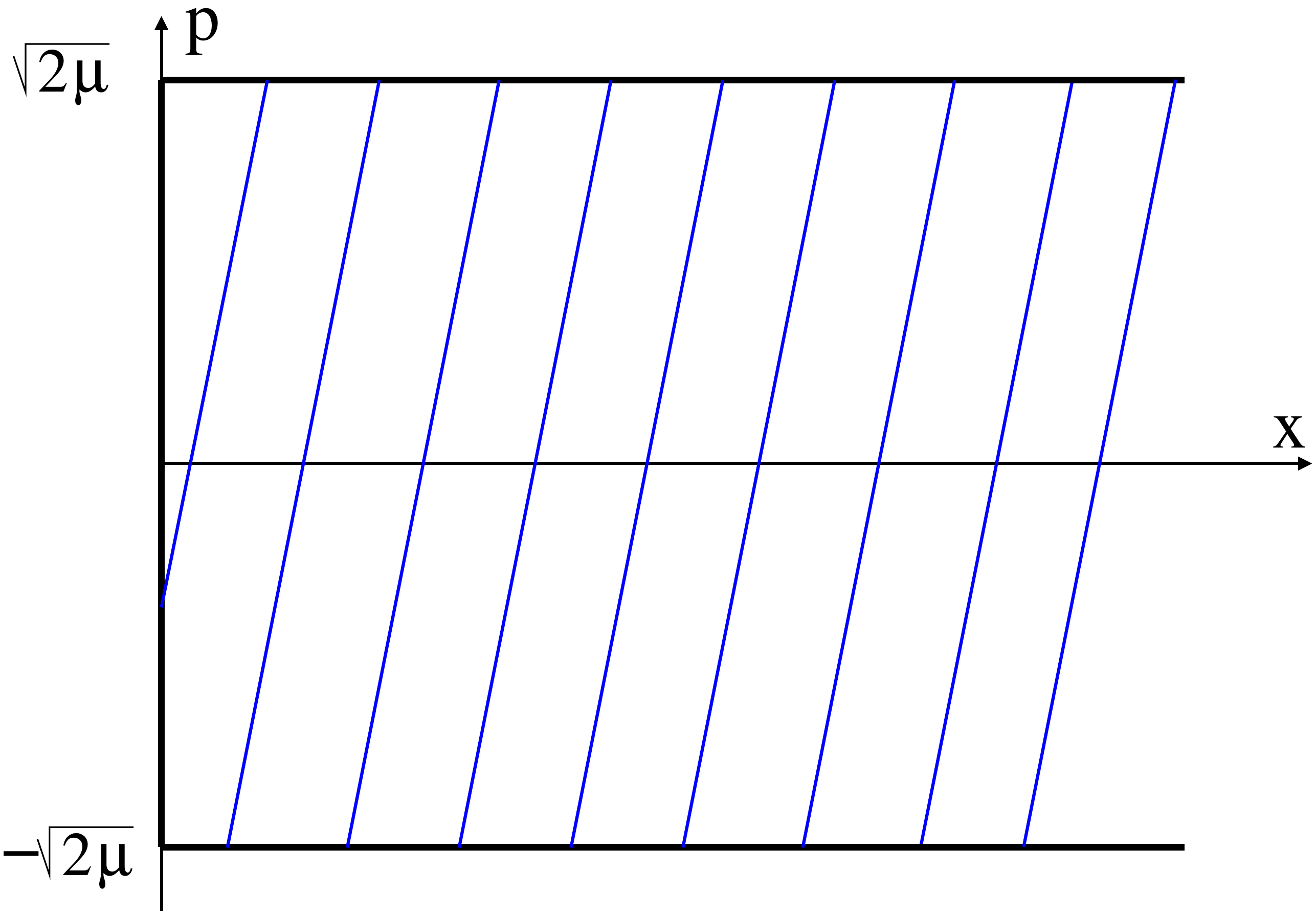}
    \caption{Illustration of the LDA prediction (\ref{LDA}) in the $(x,p)$-plane for the Wigner function of the ground-state of noninteracting fermions on the semi-infinite line with a hard wall at the origin (\ref{V_flat_single_hw}). The corresponding Fermi energy is $\mu$. Inside the 
  half-strip ${\cal S}$ (blue striped area) the Wigner function is nonzero and constant, i.e. $W_\mu(x,p) = 1/(2 \pi)$ while it vanishes outside this half-strip.}
    \label{fig:fermisurf12}
  \end{center}
\end{figure}

We first start with the case of $N$ noninteracting spinless fermions in a flat potential with a single hard wall at the origin 
\bea \label{V_flat_single_hw}
V(x) = 
\begin{cases}
&+ \infty \;, \; x<0\;, \\
&0 \; , \; \hspace*{0.5cm} x > 0 \;. \\
\end{cases}
\eea
We focus on zero temperature, where the energy levels are filled
up to the Fermi energy $\mu = k_F^2/2$. For such a potential (\ref{V_flat_single_hw}), the prediction from the LDA (\ref{LDA}) is simply (see Fig. \ref{fig:fermisurf12}) 
\bea \label{LDA_strip}
W_\mu(x,p) = 
\begin{cases}
&\dfrac{1}{2\pi} \quad, \quad (x,p) \in {\cal S} \\
& \\
&0 \quad, \quad \quad (x,p) \notin {\cal S}
\end{cases} \quad \quad, \quad \quad {\cal S} = \left\{(x,p) \; | \; x>0 \;\; \& \;\; -\sqrt{2\mu}< p < + \sqrt{2 \mu}\right\} \;.
\eea

It turns out that the structure of the Wigner function in this case is much richer than the one predicted by the LDA, as can be seen from an exact computation of $W_\mu(x,p)$.  We start with the exact single particle eigenfunctions given by 
\bea \label{sine}
\phi_k(x) = \sqrt{\frac{2}{\pi}} \Theta(x) \sin{(k\,x)} \;,\; k > 0 \;,
\eea
with corresponding energies $\epsilon_k = k^2/2$.
The Wigner function is obtained by inserting the explicit expression for the eigenfunctions (\ref{sine}) in the general formula given in Eq. (\ref{eq:Wa}), replacing the discrete sum over $n$ by an integral over $k$ since we have a continuous spectrum of states in this case. This yields the exact formula
\bea \label{Wigner_flat_hw}
W_\mu(x,p) = \frac{1}{\pi^2} \int_{-2x}^{2x} dy\, e^{i p y} \int_0^{k_F} dk\, \sin{\left[k\left(x - \frac{y}{2} \right)\right]} \sin{\left[k\left(x + \frac{y}{2}\right)\right]}  \;.
\eea
The integral over $k$ is easily done, leading to
\bea \label{Wigner_flat_hw_2}
W_\mu(x,p) = \frac{1}{\pi^2} \int_{-2x}^{2x} dy\, e^{i p y} \left[ \frac{\sin(k_F y)}{2y} - \frac{\sin(2 k_F x)}{4x} \right] \;.
\eea
Performing the integral over $y$ one finally obtains
\bea \label{Wigner_flat_hw_3}
W_\mu(x,p) = \frac{{\rm Si}(2x(k_F+p))}{2\pi^2} +  \frac{{\rm Si}(2x(k_F-p))}{2\pi^2} - \frac{\sin{(2 k_F x) \sin{(2\, p\,x)}}}{2 \pi^2 \, p \,x} \;.
\eea
Note that this result can also be obtained by integrating over $k$ from $k=0$ to $k=k_F$ the expression for the Wigner function of a single particle with a hard wall at the origin in Eq. (\ref{W_HW}). As discussed in the Appendix \ref{app:single}, the first two sine-integral terms in (\ref{Wigner_flat_hw_3}) are reminiscent of the ``broadened'' delta-functions, this broadening being caused by the presence of the wall, while the last term comes from quantum interferences [see Eqs. (\ref{W_2PW}) and (\ref{W_HW})]. An interesting consequence of this broadening is that the Wigner function is nonzero even for $p > k_F = \sqrt{2 \mu}$, a property which is not captured by the LDA prediction (\ref{LDA_strip}).

Finally, in terms of the scaled variables $\tilde s = k_F x$ [which measures the scaled distance from the wall as in the text, see Eq. (\ref{region_III})] and $\tilde p = p/k_F$ the Wigner function in (\ref{Wigner_flat_hw_3}) reads
\bea  \label{Wigner_flat_hw_4}
W_\mu(x,p) = \frac{1}{2\pi}{\mathcal{W}}_{\rm III} \left(\tilde s = k_F x,  \tilde p = p/k_F \right) \;,
\eea
where ${\mathcal{W}}_{\rm III}(\tilde s, \tilde p)$ is the scaling function describing the region III of the hard box (see Fig. \ref{fig:fermisurf}) and is given in Eq. (\ref{region_III}). We emphasize that the result in Eq. (\ref{Wigner_flat_hw_3}) is actually exact for this model (\ref{V_flat_single_hw}). Note finally that if one sets $p =\tilde p k_F$ in the exact expression for ${\cal W}_{\rm III}(x,p)$ in Eq. (\ref{Wigner_flat_hw_3}) and then take the limit $\mu \to \infty$, or equivalently $k_F \to \infty$, one finds
\bea \label{WIIILDA}
\lim_{k_F \to \infty} {\cal W}_{\rm III}(x,p = \tilde p k_F) = \Theta(1-\tilde p) \;,
\eea
which coincides with the LDA prediction (\ref{LDA_strip}) in this limit, as expected. 

\subsection{The case of an inverse square potential}\label{app:single-hardwall-inv}


\begin{figure}[t]
  \begin{center}
    \includegraphics[width=0.4\textwidth]{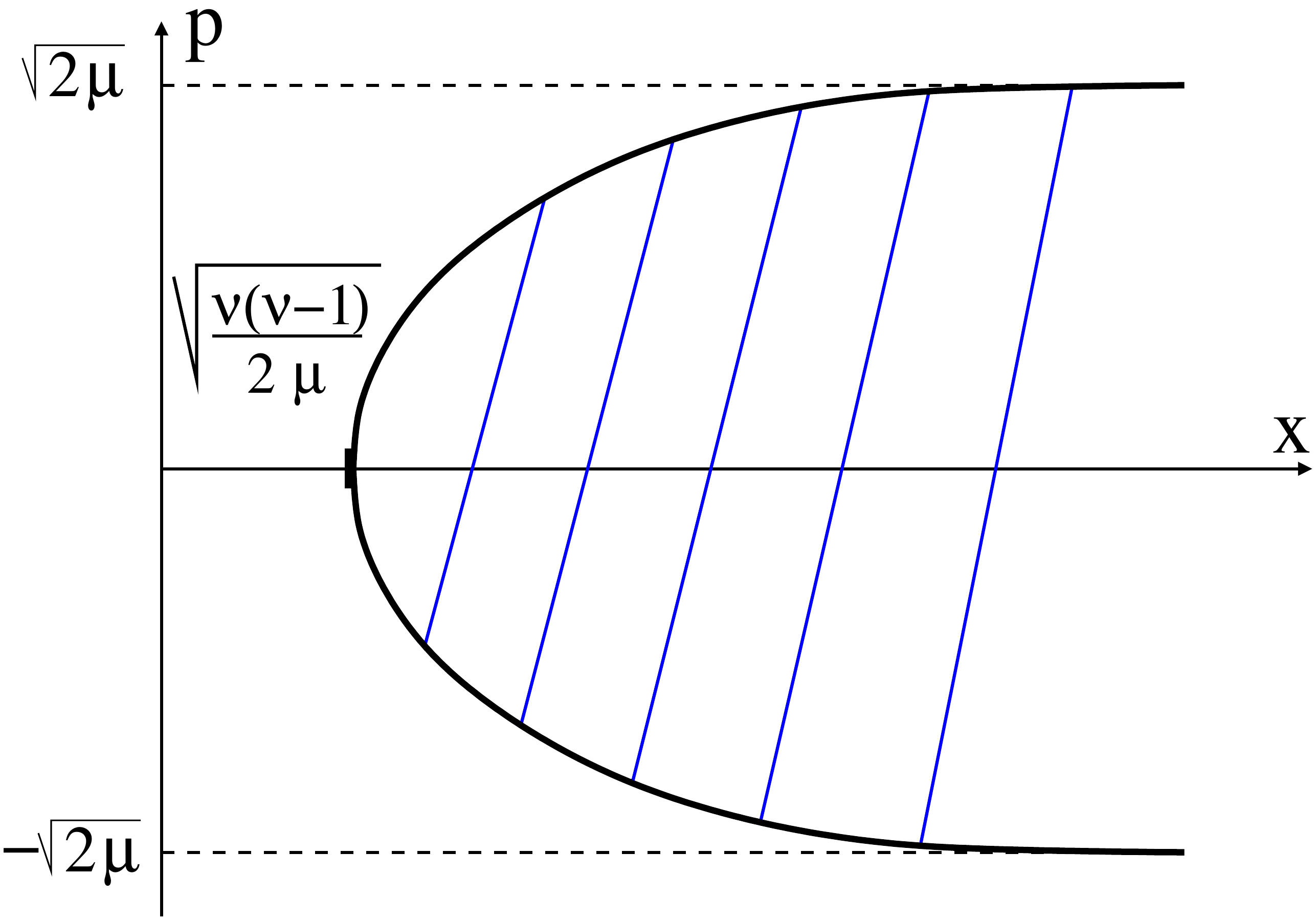}
    \caption{Illustration of the LDA prediction (\ref{LDA}) in the $(x,p)$-plane for the Wigner function of the ground-state of noninteracting fermions on the semi-infinite line with a hard wall at the origin and in the presence of an inverse square potential (\ref{V_inv_sq_single_hw}). The corresponding Fermi energy is $\mu$. Inside the 
  blue striped area the Wigner function is nonzero and constant, i.e. $W_\mu(x,p) = 1/(2 \pi)$ while it vanishes outside this half-strip. }
    \label{fig:fermisurfinv}
  \end{center}
\end{figure}

Here we consider the case of $N$ noninteracting spinless fermions in an inverse square potential and a hard wall at the origin
\bea \label{V_inv_sq_single_hw}
V(x) = 
\begin{cases}
&+ \infty \;, \; \quad \quad \;x<0\;, \\
& \\
&\dfrac{\nu(\nu-1)}{2x^2} \; , \; \hspace*{0.5cm} x > 0 \;, \\
\end{cases}
\eea
with $\nu \geq 1$. We focus on the ground state, where the energy levels are filled
up to the Fermi energy $\mu = k_F^2/2$. For such a potential (\ref{V_inv_sq_single_hw}), the prediction from the LDA (\ref{LDA}) is simply that $W_{\mu}(x,p) = 1/(2\pi)$ for $(x,p)$ inside the blue striped area shown in Fig. \ref{fig:fermisurfinv} while $W_{\mu}(x,p) = 0$ outside this region. Note that for $\mu \to \infty$, with $\nu$ fixed, this yields back the Wigner function obtained for the semi-infinite system in a flat potential and a hard wall at the origin in Eq.~(\ref{LDA_strip}). 

In this case, however, it is also possible to compute exactly the Wigner function, which displays a much richer structure than the LDA prediction. Indeed, for this potential (\ref{V_inv_sq_single_hw}) the  
single particle eigenfunctions can be computed exactly. They are given by
\bea \label{bessel}
\phi_k(x) = \sqrt{k x} \, J_{\nu-1/2}(k x) \;, \; k > 0 \;,
\eea
where $J_\nu(x)$ is the standard Bessel function of the first kind, and their corresponding energies are $\epsilon_k = k^2/2$. Note that in the case $\nu=1$, using $J_{1/2}(x) = \sqrt{2/(\pi x)} \sin(x)$, one recovers the case studied above [see Eq. (\ref{sine})]. In the ground state, the Wigner function is given by inserting the explicit expression for the eigenfunctions (\ref{bessel}) in Eq. (\ref{eq:Wa}) and by replacing the discrete sum over $n$ by an integral of $k$. This yields
\bea \label{W_bessel}
W_\mu(x,p) = \frac{1}{2\pi} \int_{-2x}^{2x} dy \, e^{i p y} \sqrt{x^2-\frac{y^2}{4}}\int_0^{k_F} dk\, k \, J_{\nu-1/2} \left(k (x-\frac{y}{2})\right) J_{\nu-1/2}\left(k(x+\frac{y}{2})\right) \;.
\eea
The integral over $k$ can be performed explicitly, yielding the result (performing also the change of variable $z=y/2$)
\bea \label{W_bessel_2}
\hspace*{-1.5cm}&&W_\mu(x,p) = \frac{k_F}{4\pi} \int_{-x}^{x} \frac{dz}{xz} \, e^{2i p z} \sqrt{x^2-z^2} \Big[(x+z) J_{\nu-1/2}(k_F(x-z)) J_{\nu+1/2}(k_F(x+z))  \nonumber \\
\hspace*{-1.5cm}&&\hspace*{6.5cm}- (x-z) J_{\nu+1/2}(k_F(x-z)) J_{\nu-1/2}(k_F(x+z)) \Big].
\eea
Performing the change of variable $z=u/k_F$ one finds that $W_\mu(x,p)$ takes the scaling form
\bea \label{scaling_W_bessel}
&&W_\mu(x,p) = \frac{1}{2\pi}{\cal W}_{\rm III, \nu}(\tilde  s = k_F x, \tilde p = p/k_F) \;,\\
&&{\cal W}_{{\rm III}, \nu}(\tilde  s,\tilde p) = \frac{1}{2 \tilde  s} \int_{- \tilde  s}^{\tilde  s} \frac{du}{u} e^{2i \tilde p u} \sqrt{\tilde  s^2 - u^2} \Big[(\tilde  s+u) J_{\nu-1/2}(\tilde  s-u) J_{\nu+1/2}(\tilde  s+u) \nonumber \\
&&\hspace*{6.5cm}- (\tilde  s-u) J_{\nu+1/2}(\tilde  s-u) J_{\nu-1/2}(\tilde  s+u) \Big] \;. \label{scaling_W_bessel_explicit}
\eea
Interestingly, we see that the Wigner function ${\cal W}_{{\rm III}, \nu}(\tilde  s,\tilde p)$ depends continuously on the parameter $\nu$. In particular, 
setting $\nu=1$ in (\ref{scaling_W_bessel_explicit}) one can check that ${\cal W}_{{\rm III},\nu = 1}(\tilde  s, \tilde p) = {\cal W}_{\rm III}(\tilde  s, \tilde p)$ given in Eq. (\ref{region_III}), as expected. Note that, for generic $\nu$, it seems difficult to evaluate the remaining integral over $u$ -- although it seems possible (though cumbersome) for $\nu = 2, 3, \ldots$. One can however easily evaluate numerically the integral in Eq. (\ref{scaling_W_bessel_explicit}) for different values of $\nu$ and generic values of $\tilde s$ and $\tilde p$. Note also that this integral representation in Eq. (\ref{scaling_W_bessel_explicit}) is in principle also amenable to a precise analysis of the various asymptotic behaviours of ${\cal W}_{{\rm III}, \nu}(\tilde  s,\tilde p)$, similar to the one carried out for ${\cal W}_{{\rm III}, \nu=1}(\tilde  s,\tilde p)={\cal W}_{{\rm III}}(\tilde  s,\tilde p)$ in the text [see Eqs. (\ref{W3_smalls})-(\ref{W3_largep})].  

Let us conclude this Section by recalling that close to the origin, the quantum correlations of the fermions in the ground state of the inverse square potential in Eq. (\ref{V_inv_sq_single_hw}) are described 
by the so-called Bessel kernel \cite{lacroix2018non}, which is well known in RMT \cite{forrester}. This kernel depends continuously on $\nu$ and, as $\nu \to \infty$, one can show (see e.g. \cite{lacroix2018non}) that the Bessel kernel, properly centered and scaled, converges to the Airy kernel, which describes the edge properties of the Fermi gas in the presence of a smooth potential \cite{fermions_review}. Therefore one expects that, in this limit $\nu \to \infty$, the limiting scaling function ${\cal W}_{{\rm III}, \nu}(\tilde  s,\tilde p)$ properly centered and scaled should converge to the scaling function ${\cal W}(a)$ in Eq. (\ref{scal2}) found for smooth potentials \cite{DDMS2018}. This family of scaling function ${\cal W}_{\rm III,\nu}(\tilde s, \tilde p)$ thus smoothly interpolates between the hard-wall scaling function ${\cal W}_{\rm III}(\tilde s, \tilde p)$ discussed in this paper in Eq. (\ref{region_III}) as $\nu \to 1$ and the one found previously for smooth potentials, i.e., ${\cal W}(a)$ in (\ref{scal2}), as $\nu \to \infty$. We have not tried, however, to study this crossover in detail.

\section{Wigner function noninteracting fermions in the presence of a single $d$-dimensional hard wall} \label{app:single-hardwall_d}
  
In this Appendix, we compute exactly the Wigner functions for $N$ noninteracting fermions in the presence of the $d$-dimensional hard-wall potential given in Eq. (\ref{hw_ddim}). In this case, the exact eigenfunctions are indexed by a vector ${\bf k} = (k_1, k_2, \cdots, k_d)$
\bea \label{ef_HWd}
\phi_{{\bf k}}({\bf x}) = \sqrt{\frac{2}{\pi}}\, \frac{1}{(\sqrt{2 \pi})^{d-1}} \, \Theta(x_d)\,\sin{(k_d x_d)} \, e^{i \sum_{j=1}^{d-1} k_j \, x_j} = \frac{1}{2^{\frac{d-2}{2}}} \frac{1}{\pi^{\frac{d-1}{2}}} \Theta(x_d)\,\sin{(k_d x_d)} e^{i {\bf k}_t \cdot {\bf x}_t }\;, {\rm with} \;\;  k_d > 0\;,
\eea
where we used the notation ${\bf x}_t = (x_1, x_2, \cdots, x_{d-1})$ and similarly ${\bf k}_t = (k_1, k_2, \cdots, k_{d-1})$. The Wigner function in the ground state of fermions with Fermi energy $\mu = \sqrt{2 k_F}$ is then given by the generalization of Eq. (\ref{Wigner_flat_hw}) to $d$ dimensions, i.e.,
\bea \label{W_HWd}
W_\mu({\bf x}, {\bf p}) = \frac{1}{(2 \pi)^d} \int d^{d-1} {\bf k}_t \int_0^\infty dk_d \, \Theta(k_F - |{\bf k}|) \int d^d {\bf y} e^{i {\bf p} \cdot {\bf y}} \phi^*_{\bf k}\left( {\bf x} + \frac{\bf y}{2}\right) \phi^*_{\bf k}\left( {\bf x} - \frac{\bf y}{2}\right) \;.
\eea
By inserting the expression for the eigenfunctions (\ref{ef_HWd}) in Eq. (\ref{W_HWd}) we see that the integrals over $y_1, y_2, \cdots, y_{d-1}$ can be performed yielding simply $(2 \pi)^{d-1} \delta({\bf k}_t - {\bf p}_t)$, where ${\bf p}_t = (p_1, p_2, \cdots, p_{d-1})$. Therefore the integrals over ${\bf k}_t$ become trivial and we get
\bea \label{W_HWd_2}
W_\mu({\bf x}, {\bf p}) = \frac{\Theta(k_F^2-{\bf p}_t^2)}{(2 \pi)^d} \frac{2}{\pi}\int_0^{\infty} dk_d \,\Theta\left(k_F - \sqrt{k_d^2 + {\bf p}_t^2}\right) \int_{-2 x_d}^{2 x_d} dy_d \, e^{i p_d y_d}  \sin\left(x_d + \frac{y_d}{2} \right)\, \sin\left(x_d - \frac{y_d}{2} \right) \;.
\eea
The remaining integrals over $k_d$ and $x_d$ are then exactly similar to the ones performed in the one-dimensional case in Eqs. (\ref{Wigner_flat_hw})- (\ref{Wigner_flat_hw_3}) with the substitutions $x \to x_d$, $p \to p_d$ and $k_F \to \sqrt{k_F^2 - {\bf p}_t^2}$. This yields
\be \label{W_HWd_2}
W_\mu({\bf x}, {\bf p})  = \frac{\Theta(k_F^2-{\bf p}_t^2)}{(2 \pi)^d} \left(\frac{{\rm Si}(2x_d( \sqrt{k_F^2 - {\bf p}_t^2}+p_d))}{\pi} +  \frac{{\rm Si}(2x_d( \sqrt{k_F^2 - {\bf p}_t^2}-p_d))}{\pi} - \frac{\sin{(2  x_d\sqrt{k_F^2 - {\bf p}_t^2}) \sin{(2\, p_d\,x_d)}}}{\pi \, p_d \,x_d} \right) \;.
\ee
In the particular case $d=3$, we recover the result of Ref. \cite{ADSR1987}. We also see that this result (\ref{W_HWd_2}) coincides exactly with the expression found in Eq. (\ref{radial4}) for 
the Wigner function for $N \gg 1$ fermions in a spherical box and near the hard wall in terms of rescaled variables, i.e. with $x_d \equiv \tilde s_n/k_F$, $p_d \equiv \tilde p_n k_F$ and ${\bf p}_t = \tilde {\bf p}_t\,k_F$.

\section{Limiting Wigner function in $d$-dimensions using  a representation of the edge kernel in terms of Bessel functions}\label{app:Bessel}

In this Appendix, we provide an alternative derivation of the limiting $d$-dimensional Wigner function near a hard-wall starting from the expression for $W_N(x,p)$ given 
in Eqs. (\ref{W_d_dim_1}) and (\ref{kernel_d_largeN}) and using a representation of the edge kernel in terms of Bessel functions obtained in Ref. \cite{lacroix2018non} -- see Eqs. (125)--(127). 
This reads 
\be \label{Bessel1}
W_N({\bf x}_w + k_F^{-1} {\bf \tilde  s},  k_F {\bf \tilde p}) \underset{N \to \infty}{\longrightarrow}\frac{1}{(2 \pi)^d} \int d^{d-1} {\bf \tilde y}_t \int_{-2 \tilde  s_n}^{2 \tilde  s_n} d\tilde y_n e^{i({{\bf \tilde p}_t \cdot {\bf \tilde y}_t} + \tilde p_n \tilde y_n)} \left[ \frac{J_{d/2}(\sqrt{{\bf \tilde y}_t^2 + \tilde y_n^2})}{(2 \pi \sqrt{{\bf \tilde y}_t^2+\tilde y_n^2})^{d/2}} - \frac{J_{d/2}(\sqrt{{\bf \tilde y}_t^2 + 4\tilde  s_n^2})}{(2 \pi \sqrt{{\bf \tilde y}_t^2+4\tilde  s_n^2})^{d/2}}\right] \;.
\ee
The $d-1$-dimensional integral over ${\bf y}_t$ can be explicitly computed since the Fourier transform of a radially symmetric function. Namely one can use the formula, for any smooth function $g(z)$
\bea \label{Hankel}
\frac{1}{(2 \pi)^d}  \int d^{d-1} {\bf \tilde y}_t \, e^{i{{\bf \tilde p}_t \cdot {\bf \tilde y}_t}} \, g(| {\bf \tilde y}_t |) = \frac{1}{(2\pi)^{\frac{d+1}{2}} \tilde p_t^{d/2-3/2}} \int_0^\infty d\tilde y_t \, \tilde y_t^{d/2-1/2} \, J_{\frac{d-3}{2}} \left( \tilde p_t \tilde y_t\right) g(\tilde y_t) \quad, \quad \tilde p_t = | {\bf \tilde p_t} | \;.
\eea 
Using this relation (\ref{Hankel}) in Eq. (\ref{Bessel1}) one obtains
\begin{align}
&&  W_N({\bf x}_w + k_F^{-1} {\bf \tilde  s},  k_F {\bf \tilde p}) \underset{N \to \infty}{\longrightarrow}\ \frac{1}{(2\pi)^{{{\frac{d+1}{2}}}} \tilde p_t^{d/2-3/2}}  \int_{-2 \tilde  s_n}^{2 \tilde  s_n} d\tilde y_n e^{i \tilde p_n \tilde y_n}\int_0^\infty d\tilde y_t \tilde y_t^{d/2-{{1/2}}} J_{d/2-3/2}(\tilde p_t \tilde y_t)  \nonumber \\
&& \times \left[ \frac{J_{d/2}(\sqrt{{\tilde y}_t^2 + \tilde y_n^2})}{(2 \pi \sqrt{{\tilde y}_t^2+\tilde y_n^2})^{d/2}} - \frac{J_{d/2}(\sqrt{{\tilde y}_t^2 + 4\tilde  s_n^2})}{(2 \pi \sqrt{{\tilde y}_t^2+4\tilde  s_n^2})^{d/2}}\right] \label{eq:dhankel}\;.
\end{align}
Using the relation \cite{PBM1986} (see relation 12, p. 217) 
 \begin{align}
 \int_0^\infty dx x^{\nu+1}J_\nu(cx)\frac{J_\mu(b\sqrt{x^2+z^2})}{(x^2+z^2)^{\mu/2}} = \Theta(b-c)\frac{c^\nu z^{1+\nu-\mu}}{b^\mu}(b^2-c^2)^{(\mu-\nu-1)/2}J_{\mu-\nu-1}(z\sqrt{b^2-c^2})
 \end{align}
 specialized to $\nu=d/2-3/2$, $c=\tilde p_t$, $\mu=d/2$, $z=y_n$, $b=1$, and then to $\nu=d/2-3/2$, $c=\tilde p_t$, $\mu=d/2$, $z=2\,s_n$, $b=1$ to evaluate the two integrals in (\ref{eq:dhankel}) one gets, using $J_{1/2}(z) = \sqrt{\pi/(2z)} \sin(z)$
\begin{align}\label{App_Bessel1}
&&  W_N({\bf x}_w + k_F^{-1} {\bf \tilde  s},  k_F {\bf \tilde p}) \underset{N \to \infty}{\longrightarrow} \frac{\Theta(1-{\bf p}_t^2)}{2^d \pi^{d+1}} \int_{-2 \tilde s_n}^{2 \tilde s_n} d \tilde y_n e^{i \tilde p_n \tilde y_n} \left[ \frac{\sin{(\tilde y_n \sqrt{1 - {\bf p}_t^2})}}{\tilde y_n} - \frac{\sin{(2\tilde s_n \sqrt{1 - {\bf p}_t^2})}}{2\tilde s_n} \right] \;.
\end{align} 
Finally, performing the integral over $\tilde y_n$ one arrives at the expression given in in Eq. (\ref{radial4}) obtained in the text by a different method.

\end{appendix}

\end{document}